\documentclass[journal]{IEEEtran}
\usepackage{amsmath}
\usepackage{amsfonts}
\usepackage{amssymb}
\usepackage{amsbsy}
\usepackage{graphicx}
\usepackage{times}
\usepackage{rotating}
\usepackage{bm}
\usepackage{default}
\usepackage{color}
\usepackage{cite}

\newtheorem{algorithm}{Algorithm}
\newtheorem{theorem}{Theorem}

\newtheorem{lemma}{Lemma}

\newtheorem{example}{Example}[section]
\newtheorem{remark}{Remark}[section]

\def\I{\mathcal{I}}
\def\P{\mathcal{P}}
\def\D{\mathcal{D}}

\def\E{\mathcal{E}}
\def\A{\mathcal{A}}
\def\N{\mathcal{N}}
\def\M{\mathcal{N}}
\def\R{\mathcal{R}}
\def\AR{\mathcal{A}_R}
\def\IRR{\mathcal{I}_R}

\def\IND{\operatorname{I}}

\def\IR{\mathbb{R}}
\def\IC{\mathbb{C}}

\def\IZ{\mathbb{Z}}

\def\dt{\tilde z}
\def\e{z}
\def\d{z}

\def\mt{\widetilde m}
\def\nt{\widetilde n}

\def\m{m}
\def\n{n}

\def\muv{\boldsymbol{\mu}}

\def\kv{\mathbf{c}}
\def\xv{\mathbf{x}}

\def\uv{\mathbf{u}}
\def\hv{\mathbf{h}}
\def\xvR{\mathbf{x}_R}
\def\zv{\mathbf{z}}

\def\hv{\mathbf{h}}

\def\yv{\mathbf{y}}
\def\xva{\mathbf{x}_a}

\def\xvba{\mathbf{\bar x}_a}
\def\xvb{\mathbf{x}_b}

\def\Lam{\mathbf{\Lambda}}

\def\Cm{\mathbf{C}}
\def\Km{\mathbf{C}}
\def\Xm{\mathbf{X}}
\def\Ym{\mathbf{Y}}
\def\Zm{\mathbf{Z}}
\def\Hm{\mathbf{H}}

\def\Um{\mathbf{U}}

\newcommand{\argmin}[1]
{\underset{#1}{\operatorname{argmin}}}

\newcommand{\argmax}[1]
{\underset{#1}{\operatorname{argmax}}}

\begin{document}

\IEEEoverridecommandlockouts

\title{Merging Belief Propagation and the Mean Field Approximation: A Free Energy Approach}

\author{
Erwin Riegler,~\IEEEmembership{Member,~IEEE}, Gunvor Elisabeth Kirkelund, Carles Navarro Manch\'on, Mihai-Alin Badiu, and Bernard Henri Fleury,~\IEEEmembership{Senior Member,~IEEE}
\thanks{

Erwin Riegler is with the Institute of Telecommunications, Vienna University of Technology, Austria (e-mail: erwin.riegler@nt.tuwien.ac.at).

Gunvor Elisabeth Kirkelund, Carles Navarro Manch\'on, and Bernard Henri Fleury are with the Department of Electronic Systems, Aalborg University, Denmark 
(e-mail: \{gunvor, cnm, bfl\}@es.aau.dk).

Mihai-Alin Badiu is with the Communications Department, Technical University of Cluj-Napoca, Romania (e-mail: Mihai.Badiu@com.utcluj.ro). The research of Mihai-Alin Badiu was carried out while he visited Aalborg University.

This work was supported by the WWTF grant ICT10-066, the FWF grant
S10603-N13 within the National Research Network SISE, Renesas Mobile Corporation, 
the 4GMCT cooperative research project funded by Intel Mobile Communications, Agilent Technologies, Aalborg University and the Danish National Advanced Technology Foundation, the project ICT-248894 Wireless Hybrid Enhanced Mobile Radio Estimators 2 (WHERE2),
 and 
the project SIDOC under contract no. POSDRU/88/1.5/S/60078.  

The results of this paper have been presented partially in \cite{rikimafl10}.
}
}



\maketitle


\begin{abstract}
We present a joint message passing approach that combines belief propagation and the mean field approximation. Our analysis is based on the region-based free energy approximation method proposed by Yedidia et al. We show that the message passing fixed-point equations  obtained with this combination correspond to stationary points of a constrained  region-based free energy approximation. Moreover, we present a convergent implementation of these message passing fixed-point equations provided that the underlying factor graph fulfills certain technical conditions. In addition, we show how to include hard constraints in the part of the factor graph corresponding to belief propagation. Finally, we demonstrate an application of our method to iterative channel estimation and decoding in an orthogonal frequency division multiplexing (OFDM) system.
\end{abstract}


\section{Introduction}\label{introduction}
Variational techniques have been used for decades
in quantum and statistical physics, where they are referred to as the
{\em mean field} (MF) approximation \cite{pa88}. Later, they found their way to the area of machine learning or statistical inference, see, e.g., \cite{ghbe00, joghjasa99, CWi05, bi06}. The basic idea of variational inference is to derive the statistics of ``hidden" random variables given the
knowledge of ``visible" random variables of a certain probability density function
(pdf). 
In the MF approximation, this pdf is approximated 
 by some ``simpler," e.g.,
(fully) factorized pdf and the Kullback-Leibler divergence between the approximating and the true pdf is minimized, which can be done in an iterative, i.e., message passing like way. Apart from being fully factorized, the approximating pdf typically fulfills additional constraints that allow for messages with a simple structure, which can be updated in a simple way. For example, additional exponential conjugacy constraints result in messages propagating along the edges of the underlying Bayesian network that are described by a small number of parameters
 \cite{CWi05}. Variational inference methods were recently applied in \cite{kimachrifl10} to the
{\em channel state estimation/interference cancellation part} of a class of MIMO-OFDM receivers that iterate between detection, channel estimation, and decoding.

An approach different from the MF approximation is {\em belief propagation} (BP) \cite{JPe65}. Roughly speaking, with BP one tries to find
{\em local} approximations, which are---exactly or approximately---the
marginals of a certain pdf\footnote{Following the convention used in \cite{yefrwe04}, we use the name BP also for loopy BP.}. This can also be done in an iterative way, where messages are passed along the edges of a factor graph \cite{ksbrlo01}.
A typical application of BP is {\em decoding} of turbo  or low density parity check (LDPC) codes. Based on the excellent performance of BP, a lot of variations have been derived in order to improve the performance of this algorithm even further. For example, minimizing an upper bound on the log partition function of a pdf leads to the powerful tree reweighted BP algorithm \cite{wajawi05}.
An offspring of this
idea is the recently developed uniformly tree reweighted BP algorithm \cite{wypesa11}. Another example is \cite{iktaam04}, where methods from information geometry are used to compute correction terms for the beliefs obtained by loopy BP. An alternative approach for turbo decoding that uses projections (that are dual in the sense of \cite[Ch. 3]{Amari} to the one used in \cite{iktaam04}) on constraint subsets can be found in \cite{mogu98}. A combination of the approaches used in\cite{iktaam04} and in  \cite{mogu98} can be found in \cite{wajore05}.

Both methods, BP and the MF approximation, have their own virtues and disadvantages. For example, the MF approximation
\begin{description}
\item[+] \hspace*{-5truemm}always admits a convergent implementation;
\item[+] \hspace*{-5truemm}has simple message passing update rules, in particular\\\hspace*{-5truemm}\noindent for conjugate-exponential models;
\item[--] \hspace*{-5truemm}is not compatible with hard constraints,
\end{description}
and  BP
\begin{description}
\item[+]\hspace*{-5truemm}yields a good approximation of the marginal\\\hspace*{-5truemm}\noindent distributions if the factor graph has no short cycles;
\item[+]\hspace*{-5truemm}is compatible with hard constraints like, e.g.,\\\hspace*{-5truemm}\noindent code constraints;
\item[--] \hspace*{-5truemm}may have a high complexity, especially when applied\\\hspace*{-5truemm}\noindent to probabilistic models involving both, discrete and\\\hspace*{-5truemm}\noindent continuous random variables.
\end{description}
Hence, it is of great benefit to apply BP and the MF approximation on the same factor graph in such a combination that their respective virtues can be exploited while circumventing their drawbacks. 
To this end, a {\em unified message passing algorithm} is needed that allows for combining both approaches.

The fixed-point equations of both BP and the MF approximation can be obtained by minimizing an approximation of the Kullback-Leibler divergence, called region-based  free energy approximation
. This approach differs from other methods, see, e.g.,
\cite{TMi05}\footnote{
An information geometric interpretation of the different objective functions used in  \cite{TMi05} can be found in \cite[Ch. 2]{Amari}.}, because the starting point for the derivation of the corresponding message passing fixed-point equations is the same objective function for both, BP and the MF approximation. 
The main technical result of our work is Theorem \ref{Theoremcombined}, where we show that the message passing fixed-point equations for such a combination of BP and the MF approximation correspond to stationary points of one single constrained region-based free energy approximation and provide a clear rule stating how to couple the messages propagating in the BP and MF part. In fact, based on the factor graph corresponding to a factorization  of a probability mass function (pmf) and a choice for a   separation of this factorization into BP and MF factors, Theorem  \ref{Theoremcombined} gives the message passing fixed-point equations for the factor graph representing the whole factorization of the pmf. One example of an application of Theorem  \ref{Theoremcombined} is joint channel estimation, interference cancellation, and decoding. Typically, these tasks are considered separately and the coupling between them is described in a heuristic way. As an example of this problematic, there has been a debate in the research community on whether a posteriori probabilities (APP) or extrinsic values should be fed back from the decoder to the rest of the receiver components; several authors coincide in proposing the use of extrinsic values for MIMO detection \cite{wapo99,weme06,romu08} while using APP
values for channel estimation \cite{weme06, romu08}, but no thorough justification for this choice is
given apart from the achieved superior performance shown by simulation results. Despite having a clear rule to update the messages for the whole factor graph representing a factorization of a pmf, an
additional advantage is the fact that solutions of fixed-point equations for the messages are related to the stationary points of the corresponding  constrained region-based free energy approximation. This correspondence is important because it yields an interpretation of the computed beliefs for arbitrary factor graphs similar to the case of solely BP, where solutions of the message passing fixed-point equations do in general not correspond to the true marginals if the factor graph has cycles but always correspond to stationary points of the constrained Bethe free energy \cite{yefrwe04}. Moreover, this observation allows us to present  a systematic way of updating the messages, namely, Algorithm \ref{Algorithm}, that is guaranteed to converge provided that the factor graph representing the  factorization of the pmf fulfills certain technical conditions.

The paper is organized as follows. In the remainder of this section we fix our notation. Section \ref{Knownresults} is devoted to the introduction of the region-based free energy approximations proposed by \cite{yefrwe04} and to recall how BP, the MF approximation, and the EM algorithm \cite{delaru77} can be obtained by this method.
Since the MF approximation is typically used for parameter estimation, we briefly show how to extend it to the case of continuous random variables using an approach presented already in \cite[pp. 36--38]{ku78} that avoids complicated methods from variational calculus.
Section \ref{secC} is the main part of this work. There we state our main result, namely, Theorem  \ref{Theoremcombined}, and show how the message passing fixed-point equations of a combination of BP and the MF approximation can be related to the stationary points of the corresponding constrained region-based free energy approximation.
We then  (i) prove Lemma \ref{theohard}, which generalizes Theorem \ref{Theoremcombined} to the case where the factors of the pmf in the BP part are no longer restricted to be strictly positive real-valued functions, and (ii) present Algorithm \ref{Algorithm} that is a convergent implementation of the message passing update equations presented in Theorem \ref{Theoremcombined} provided that the factor graph representing the factorization of the pmf fulfills certain technical conditions. As a byproduct, (i) gives insights into solely BP (which is a special case of the combination of BP and the MF approximation) with hard constraints, where only conjectures are formulated in \cite{yefrwe04}.  In Section \ref{Application} we apply  Algorithm \ref{Algorithm} to joint channel estimation and decoding in an OFDM system. More advanced receiver architectures together with numerical simulations and a comparison with other state of the art receivers 
can be found in \cite{makirichfl11} and an additional application of the algorithm in a cooperative communications scenario is presented in \cite{bamabofl12}. 
Finally, we conclude in Section \ref{Conclusion} and present an outlook for further research directions.


\subsection{Notation}
Capital calligraphic letters $\A, \I, \N$ denote finite sets. The cardinality of a set $\I$ is denoted by
$|\I|$.  If $i\in\I$ we write $\I\setminus i$ for $\I\setminus \{i\}$.
We use the convention that
$\prod_{\emptyset}(\dots)\triangleq 1$, where $\emptyset$
denotes the empty set. For any finite set $\I$, $I_\I$ denotes the indicator function on $\I$, i.e., $\IND_\I(i)=1$ if $i\in\I$
and $\IND_\I(i)=0$ else. We denote by capital letters $X$ discrete random variables with a finite number of realizations and pmf $p_X$. For a random variable $X$, we use the convention that $x$ is a representative for all possible realizations of $X$, i.e., $x$ serves as a running variable, and denote a particular realization by $\bar x$. For example,
$\sum_x(\dots)$ runs through all possible realizations $x$ of $X$ and for two functions $f$ and $g$ depending on all realizations $x$ of $X$, $f(x)=g(x)$  means that $f(\bar x)=g(\bar x)$ for each particular realization $\bar x$ of $X$. If $F$ is a functional of a pmf $p_X$ of a random variable $X$ and $g$ is a function depending on all realizations $x$ of X, then
$\frac{\partial F}{\partial p(x)}=g(x)$ means that  $\frac{\partial F}{\partial p(\bar x)}=g(\bar x)$ is well defined and holds for each particular realization
$\bar x$ of $X$. We write $\xv=(x_i\mid i\in\I)^{\operatorname{T}}$ for the realizations of the vector of random variables $\Xm=(X_i\mid i\in\I)^{\operatorname{T}}$. If $i\in\I$, then $\sum_{\xv\setminus x_i}(\dots)$ runs through all possible realizations of
$\Xm$ but $X_i$.
For any nonnegative real valued function $f$ with argument $\xv=(x_i\mid i\in\I)^{\operatorname{T}}$ and $i\in\I$, $f\mid_{\bar x_i}$ denotes
$f$ with fixed argument $x_i=\bar x_i$. If a function $f$ is identically zero, we write $f\equiv 0$ and $f\not\equiv 0$ means that
it is not identically zero. For two real valued functions $f$ and $g$ with the same domain and argument $x$, we write $f(x)\propto g(x)$ if $f=c g$ for some real positive constant $c\in\IR_+$.
We use the convention that $0\ln(0)=0$, $a\ln(\frac{a}{0})=\infty$ if $a>0$, and
$0\ln(\frac{0}{0})=0$ \cite[p.\,31]{Cover91}. For $x\in\IR$, $\delta(x)=1$ if $x=0$ and zero else.
Matrices 
are denoted by capital boldface Greek letters.
The superscripts ${}^{\operatorname{T}}$ and ${}^{\operatorname{H}}$ stand for transposition and
Hermitian transposition, respectively.
For a matrix $\Lam\in\IC^{m\times n}$, the entry in the
$i$th row and $j$th column is denoted by $\lambda_{i,j}=[\Lam]_{i,j}$.
For two vectors
$\xv=(x_i\mid i\in\I)^{\operatorname{T}}$ and $\yv=(y_i\mid i\in\I)^{\operatorname{T}}$,
$\xv\odot\yv=(x_iy_i\mid i\in\I)^{\operatorname{T}}$ denotes the Hadamard product of $\xv$ and $\yv$.
Finally, $\text{CN}(\xv;\muv,\mathbf{\Sigma})$ stands for the pdf of a jointly proper complex Gaussian random vector $\Xm\sim\mathcal{CN}(\muv,\mathbf{\Sigma})$ with mean $\muv$ and covariance matrix $\mathbf{\Sigma}$.

\section{Known results}\label{Knownresults}

\subsection{Region-based free energy approximations \cite{yefrwe04}}\label{secRB}
Let $p_{\Xm}$ be a certain positive pmf of a vector $\Xm$ of random variables $X_i$ $(i\in\I)$ that factorizes as
\begin{align}\label{eq:factorization}
p_{\Xm}(\xv)=\prod_{a\in\A}f_a(\xva)
\end{align}
where
$\xv\triangleq (x_i\mid i\in\I)^{\operatorname{T}}$ and $\xv_a\triangleq (x_i\mid i\in\N(a))^{\operatorname{T}}$ with $\N(a)\subseteq\I$ for all $a\in\A$. Without loss of generality we assume that  $\A\cap\I=\emptyset$, which can always be achieved by renaming indices.\footnote{
For example, we can write
\begin{align*}
\I&=\{1,2,\dots,|\I|\}\\
\A&=\{\overline 1,\overline 2, \dots,\overline{|\A|}\}.
\end{align*}
This implies that any function that is defined pointwise on $\A$ and $\I$ is well defined. For example, if in addition to the definition of the sets $\N(a)$  ($a\in\A$) we set
$\M(i)\triangleq\{a\in\A\mid i\in\N(a)\}$ for all $i\in\I$, the function
\begin{align*}
\N: \I\cup\A &\to \Pi(\I\cup\A)\\
a&\mapsto \N(a),\quad\text{for all}\ a\in\A\\
i&\mapsto \N(i),\quad\text{\,for all}\ i\in\I
\end{align*}
with $\Pi(\I\cup\A)$ denoting the collection of all subsets of $\I\cup\A$ is well defined  because $i\neq a$ for all $i\in\I, a\in\A$.}
Since  $p_{\Xm}$ is a strictly positive pmf, we can assume without loss of generality that all the factors $f_a$ of  $p_{\Xm}$ in \eqref{eq:factorization} are real-valued positive functions. Later in Section \ref{secC}, we shall show how to relax the positivity constraint for some of these factors.
The factorization in \eqref{eq:factorization} can be visualized in a {\em factor graph} \cite{ksbrlo01}\footnote{Throughout the paper we work with Tanner factor graphs as opposed to Forney factor graphs.}.
In a factor graph, $\N(a)$ is the set of all variable nodes connected to a factor node $a\in\A$ and  $\M(i)$ represents the set of all factor nodes connected to a variable node $i\in\I$. An example of a factor graph is depicted in Figure \ref{Fig1}.

A {\em region} $R\triangleq(\IRR,\AR)$
consists of subsets of indices
$\IRR\subseteq\I$ and
$\AR\subseteq\A$
with the restriction that $a\in\AR$ implies that
$\N(a)\subseteq\IRR$.
To each region $R$ we associate a {\em counting number} $c_{R}\in\IZ$. A set $\R\triangleq\{(R,c_{R})\}$ of regions and associated counting numbers is called {\em valid} if
\begin{align*}
 &\sum_{(R,c_{R})\in\R} c_R \IND_{\AR}(a) =\!\!\!\! \sum_{(R,c_{R})\in\R} c_R \IND_{\IRR}(i) =1
\end{align*}
for all $a\in\A, i\in\I$.

For a positive function $b$ approximating   $p_{\Xm}$, 
we define the {\em{variational free energy}} \cite{yefrwe04}\footnote{If $p_{\Xm}$ is not normalized to one, the definition of the variational free energy contains an additional normalization constant, called Helmholtz free energy \cite[pp. 4--5]{yefrwe04}.}
\begin{equation}\label{free}
\begin{split}
F(b)
&\triangleq \sum_\xv b(\xv)\ln\frac{b(\xv)}{p_{\Xm}(\xv)}\\
&=\underbrace{\sum_\xv b(\xv)\ln b(\xv)}_{\triangleq -H(b)} - \underbrace{\sum_\xv b(\xv)\ln p_{\Xm}(\xv)}_{\triangleq -U(b)}.
\end{split}
\end{equation}
In  \eqref{free}, $H(b)$ denotes the entropy \cite[p.\,5]{Cover91} of $b$ and $U(b)$ is called  average
energy of $b$. Note that $F(b)$ is the Kullback-Leibler divergence \cite[p.\,19]{Cover91} between
$b$ and $p_{\Xm}$, i.e., $F(b)=D(b\mid\mid p_{\Xm})$. For a set $\R$ of regions and associated counting numbers,
the {\em region-based free energy approximation} is defined as \cite{yefrwe04} $F_\R\triangleq U_\R-H_\R$ with
\begin{align*}
U_\R&\triangleq-\sum_{(R,c_R)\in\R}c_R\sum_{a\in\AR}\sum_{\xvR}b_R(\xvR)\ln f_a(\xva)\\
H_\R&\triangleq-\sum_{(R,c_R)\in\R}c_R\sum_{\xvR} b_R(\xvR)\ln b_R(\xvR).
\end{align*}
Here, each $b_R$ 
is defined locally on a region $R$. Instead of minimizing
$F$ with respect to $b$, we minimize $F_\R$ with respect to all $b_R$ $((R,c_R)\in\R)$, where
the $b_R$ have to fulfill certain constraints. The quantities $b_R$ are called {\em beliefs}.
We give two examples of valid sets of regions and associated counting numbers.

\begin{example}\label{example1}
The trivial example $\R_{\text{MF}}\triangleq\{((\I,\A),1)\}$. It
leads to the MF
fixed-point equations, as will be shown in Subsection \ref{secMF}.
\end{example}

\begin{example}\label{example2}  \label{bethe}
We define two types of regions:
\begin{enumerate}
 \item {\em large regions:}  $R_a\triangleq (\N(a),\{a\})$, with $c_{R_a}=1$ for all $a\in\A$;
 \item {\em small regions:}  $R_i\triangleq(\{i\},\emptyset)$, with $c_{R_i}=1-|\M(i)|$ for all $i\in\I$.
\end{enumerate}
Note that this definition is well defined due to our assumption that $\A\cap\I=\emptyset$.
The region-based free energy approximation corresponding to the valid set of regions and associated counting numbers
\begin{align*}
\R_{\text{BP}}\triangleq\{(R_i,c_{R_i})\mid i\in\I\}\cup\{(R_a,c_{R_a})\mid a\in\A\}
\end{align*}
is called the {\em Bethe free energy} \cite{be35,yefrwe04}. It
leads to the BP
fixed-point equations, as will be shown in Subsection \ref{secBP}.
The Bethe free energy is equal to the variational free energy when the factor graph has no cycles \cite{yefrwe04}.
\end{example}

\subsection{BP fixed-point equations}\label{secBP}
The fixed-point equations for BP can be obtained from the Bethe free energy
by imposing additional marginalization and normalization constraints and computing the stationary points of the corresponding Lagrangian function\cite{he03,yefrwe04}.
The Bethe free energy reads
\begin{align}\label{bethefree}
 F_{\text{BP}}=&
\sum_{a\in\A}\sum_{\xva}b_a(\xva)\ln\frac{b_a(\xva)}{f_a(\xva)}\nonumber\\
&-\sum_{i\in\I}(|\M(i)|-1)\sum_{x_i}b_i(x_i)\ln b_i(x_i)
\end{align}
with
$b_a\triangleq b_{R_a}$ for all $a\in\A$,
$b_i\triangleq b_{R_i}$ for all $i\in\I$,
and $F_{\text{BP}}\triangleq F_{\R_\text{BP}}$.
The normalization constraints for the beliefs $b_a$ $(a\in\A)$ and the marginalization constraints for the beliefs $b_a$ and $b_i$ $(a\in\A, i\in\N(a))$
can be included in the Lagrangian  \cite[Sec.\,3.1.3]{be03}
\begin{align}\label{LBethe}
 L_\text{BP}\triangleq &\, F_\text{BP}-\sum_{a\in\A}\sum_{i\in \N(a)}\sum_{x_i}\lambda_{a,i}(x_i)
\Big(b_i(x_i)-\sum_{\xva\setminus x_i}b_a(\xva)\Big)\nonumber\\
&\, -\sum_{a\in\A}\gamma_a\Big( \sum_{\xva}b_a(\xva)-1\Big).
\end{align}
The stationary points of the Lagrangian in \eqref{LBethe} are then related to the BP fixed-point equations by the following theorem.

\begin{theorem}\cite[Th.\,2]{yefrwe04}\label{the1}
Stationary points of the Lagrangian in  \eqref{LBethe} must be BP fixed-points with
positive beliefs fulfilling
\begin{equation}\label{stationarybethe}
\begin{cases}
\begin{split}
 b_a(\xv_a)&=\, \d_a\, f_a(\xva)\prod_{i\in \N(a)}\n_{i\to a}(x_i),\quad\text{for all}\  a\in\A\\
 b_i(x_i)  &=\, \prod_{a\in \M(i)}\m_{a\to i}(x_i),\quad\text{for all}\ i\in\I
\end{split}
\end{cases}
\end{equation}
with
\begin{equation}\label{eq:fp2}
\begin{cases}
\begin{split}
\m_{a\to i}(x_i)&=\d_a \sum_{\xva\setminus x_i} f_a(\xva)\prod_{j\in \N(a)\setminus i}\n_{j\to a}(x_j)\\
\n_{i\to a}(x_i)&= \prod_{c\in \M(i)\setminus a}\m_{c\to i}(x_i)
\end{split}
\end{cases}
\end{equation}
for all $a\in\A, i\in\N(a)$ and vice versa.
Here, $\d_a$ $(a\in\A)$ are positive constants that ensure that the beliefs $b_a$ $(a\in\A)$ are normalized to one.
\end{theorem}

Often, the following alternative system of fixed-point equations is solved instead of \eqref{eq:fp2}.
\begin{equation}\label{eq:fp2a}
\begin{cases}
\begin{split}
\mt_{a\to i}(x_i)&=\omega_{a,i}\sum_{\xva\setminus x_i} f_a(\xva)\prod_{j\in \N(a)\setminus i}\nt_{j\to a}(x_j)\\
\nt_{i\to a}(x_i)&= \prod_{c\in \M(i)\setminus a}\mt_{c\to i}(x_i)
\end{split}
\end{cases}
\end{equation}
for all $a\in\A, i\in\N(a)$, 
where $\omega_{a,i}$ $(a\in\A, i\in\N(a))$ are arbitrary positive constants.
The reason for this is that for a fixed scheduling the messages computed in \eqref{eq:fp2} differ from the messages computed in \eqref{eq:fp2a} only by positive constants, which drop out when the beliefs are normalized. See also \cite[Eq. (68) and Eq. (69)]{yefrwe04}, where the $``\propto"$ symbol is used in the update equations indicating that the normalization constants are irrelevant. A solution of  \eqref{eq:fp2a} can be obtained, e.g., by updating corresponding likelihood ratios of the messages in \eqref{eq:fp2} or by updating the messages according to \eqref{eq:fp2} but ignoring the normalization constants $\d_a$ $(a\in\A)$. The algorithm converges if the normalized beliefs do not change any more. Therefore, a rescaling of the messages is irrelevant and a solution of \eqref{eq:fp2a} is obtained.
However, we note that a rescaled solution of \eqref{eq:fp2a} is not necessarily a solution of \eqref{eq:fp2}. Hence, the beliefs obtained by solving \eqref{eq:fp2a} need not be stationary points of the  Lagrangian in \eqref{LBethe}. To the best of our knowledge, this elementary insight is not published yet in the literature and  we state a necessary and sufficient condition when a solution of \eqref{eq:fp2a} can be rescaled to a solution of \eqref{eq:fp2} in the following lemma.

\begin{lemma}\label{Lemmabetter}
Suppose that $\{\mt_{a\to i}(x_i), \nt_{i\to a}(x_i)\}$ $(a\in\A, i\in\N(a))$
is a solution of \eqref{eq:fp2a} and set
\begin{align}\label{eq:datilde}
\dt_a \triangleq \frac{1}{\sum\limits_{\xva} f_a(\xva)\prod\limits_{i\in \N(a)}\nt_{i\to a}(x_i)},\quad\text{for all}\ a\in\A.
\end{align}
Then this solution can be rescaled to a solution of \eqref{eq:fp2} if and only if there exist positive constants  $g_i$ $(i\in\I)$ such that
\begin{align}\label{eq:condbetter}
\omega_{a,i}=g_i\dt_a,\quad\text{for all}\ a\in\A, i\in\N(a).
\end{align}
\end{lemma}
\proof See Appendix \ref{ProofLemmabetter}.\endproof
\begin{remark}
Note that for factor graphs that have a tree-structure the messages obtained by running the forward-backward  algorithm  \cite{ksbrlo01} always
fulfill \eqref{eq:condbetter} because we have $\omega_{a,i}=1$ $(a\in\A, i\in\N(a))$ and $\dt_a=1$ $(a\in\A)$ in this case.
\end{remark}

\subsection{Fixed-point equations for the MF approximation}\label{secMF}
A message passing interpretation of the MF approximation was derived in \cite{CWi05,da07}. In this section, we
briefly show how the corresponding fixed-point equations can be obtained by the free energy approach.
To this end, we use $\R_{\text{MF}}$ from Example \ref{example1} together with the factorization constraint\footnote{For binary random variables with pmf in an exponential family it was shown in \cite{ta00} that this gives a good approximation whenever the truncation of the Plefka expansion does not introduce a significant error.}
\begin{align}\label{eq:factorizationconstraint}
b(\xv)=\prod_{i\in\I}b_i(x_i).
\end{align}
Plugging \eqref{eq:factorizationconstraint} into the expression for the region-based free energy approximation corresponding to the trivial approximation
$\R_{\text{MF}}$ we get
\begin{equation}\label{eq:FMF}
F_\text{MF}=\sum_{i\in\I}\sum_{x_i}b_i(x_i)\ln b_i(x_i)-\sum_{a\in\A}\sum_{\xva}\prod_{i\in \N(a)} b_i(x_i)\ln f_a(\xva)
\end{equation}
with $F_\text{MF}\triangleq F_{\R_\text{MF}}$. Assuming that all the beliefs $b_i$ $(i\in\I)$ have to fulfill a normalization constraint,
the stationary points of the corresponding Lagrangian for the MF approximation can easily be evaluated to be
\begin{align}\label{eq:updateMF}
 b_i(x_i)  &= \d_i\exp
\Biggl(\
\sum_{a\in \M(i)}\sum_{\xva\setminus x_i}\prod_{j\in \N(a)\setminus i} b_j(x_j) \ln f_a(\xva)
\Biggr)
\end{align}
for all $i\in\I$, 
where the positive constants $\d_i$ $(i\in\I)$ are such that $b_i$ is normalized to one for all $i\in\I$.\footnote{
The Lagrange multiplier \cite[p.\,283]{be03} for each belief $b_i$ $(i\in\I)$ corresponding to the normalization constraint can be absorbed into
the positive constant $z_i$ $(i\in\I)$.}

For the MF approximation there always exists  a convergent  algorithm that computes beliefs $b_i$ $(i\in\I)$ solving \eqref{eq:updateMF} by simply using  \eqref{eq:updateMF}  as an iterative update equation for the beliefs. Since for all $i\in\I$
\[
\frac{\partial^2 F_\text{MF}}{\partial b_i(x_i)^2} =\frac{1}{b_i(x_i)} > 0
\]
and the set of all beliefs $b_i$ satisfying the normalization constraint $\sum_{x_i}b_i(x_i)=1$ is a convex set, 
the objective function $F_{\text{MF}}$ in \eqref{eq:FMF} cannot increase and the algorithm  is guaranteed to converge. Note that in order to derive a particular update $b_i$  $(i\in\I)$ we need all previous updates $b_j$ with
$
j\in\bigcup_{a\in \M(i)}\N(a)\setminus i.
$

By setting $n_{i\to a}(x_i)\triangleq b_i(x_i)$ for all $i\in \I, a\in \M(i)$, the fixed-point equations in \eqref{eq:updateMF} are transformed into the message passing fixed-point equations
\begin{equation}\label{eq:FPEM}
\begin{cases}
\begin{split}
n_{i\to a}(x_i)=&\, \d_i\prod\limits_{a\in \M(i)} m_{a\to i}(x_i)\\
m_{a\to i} (x_i)=&\, 
\exp\Biggl(\ \sum\limits_{\xva\setminus x_i}\prod\limits_{j\in \N(a)\setminus i}\!\!\!\! n_{j\to a}(x_j)\ln f_a(\xva)\Biggr)
\end{split}
\end{cases}
\end{equation}
for all $a\in\A, i\in\N(a)$.
The MF approximation can be extended to the case where $p_{\Xm}$ is a pdf, as shown in Appendix \ref{AppcontMF}.
Formally, each sum over $x_k$ ($k\in\I$) in \eqref{eq:updateMF} and \eqref{eq:FPEM} has to be replaced by a Lebesgue integral whenever the corresponding random variable $X_k$ is continuous.

\subsection{Expectation maximization (EM)}
Message passing interpretations for EM \cite{delaru77} were derived in \cite{ec05,dakolo05}. It can be shown that EM is a special instance of the MF approximation \cite[Sec.\,2.3.1]{Hu10}, which can be summarized as follows. 
Suppose that we apply the MF approximation to
$p_{\Xm}$ in \eqref{eq:factorization} as described before. In addition, we assume that for all $i\in\E\subseteq\I$ the beliefs $b_i$ fulfill the constraints that $b_i(x_i)=\delta(x_i-\tilde x_i)$. Using the fact that $0\ln(0)=0$, we can rewrite $F_\text{MF}$ in \eqref{eq:FMF} as
\begin{equation}\label{eq:FEM}
\begin{split}
F_\text{MF}=
&\sum_{i\in\I\setminus\E}\sum_{x_i}b_i(x_i)\ln b_i(x_i)\\
&-\sum_{a\in\A}\sum_{\xva}\prod_{i\in \N(a)} b_i(x_i)\ln f_a(\xva).
\end{split}
\end{equation}
For all $i\in\I\setminus\E$ the stationary points of $F_\text{MF}$ in \eqref{eq:FEM} have the same analytical expression as the one obtained in
\eqref{eq:updateMF}. For $i\in\E$, minimizing $F_\text{MF}$ in \eqref{eq:FEM} with respect to $\tilde x_i$ yields
\begin{align*}
\tilde x_i
&=\argmin{x_i}(F_\text{MF})\\
&=\argmax{x_i}
\Biggl( \prod_{a\in\M(i)}
\exp\Biggl(\ \sum\limits_{\xva\setminus x_i}\prod\limits_{j\in \N(a)\setminus i}\!\!\!\! b_{j}(x_j)\ln f_a(\xva)\Biggr)
\Biggr).
\end{align*}
Setting $n_{i\to a}(x_i)\triangleq b_i(x_i)$ for all $i\in \I, a\in \M(i)$, we get the message passing update equations defined in
\eqref{eq:FPEM} {\em except} that we have to replace the messages $n_{i\to a}(x_i)$ for all $i\in\E$ and $a\in\M(i)$ by
\begin{align*}
n_{i\to a}(x_i)&=\delta(x_i-\tilde x_i)
\end{align*}
with
\begin{align*}
\tilde x_i=\argmax{x_i}\Biggl(\prod\limits_{a\in \M(i)} m_{a\to i}(x_i)\Biggr)
\end{align*}
for all $i\in\E, a\in\N(a)$.
\section{Combined BP / MF approximation fixed-point equations}\label{secC}
Let
\begin{align}\label{eq:factorcombined}
p_{\Xm}(\xv)=\prod_{a\in \A_{\text{MF}}}f_a(\xva) \prod_{b\in \A_{\text{BP}}}f_b(\xvb)
\end{align}
be a partially factorized pmf with $\A_{\text{MF}}\cap \A_{\text{BP}}=\emptyset$ and
$\A\triangleq\A_{\text{MF}}\cup \A_{\text{BP}}$.
As before, we have
$\xv\triangleq (x_i\mid i\in\I)$,
$\xv_a\triangleq (x_i\mid i\in\N(a))^{\operatorname{T}}$, with $\N(a)\subseteq\I$ for all $a\in\A$, and
$\M(i)\triangleq\{a\in\A\mid i\in\N(a)\}$ for all $i\in\I$.
We refer to the factor graph representing the factorization $\prod_{a\in \A_{\text{BP}}}f_a(\xva)$ in \eqref{eq:factorcombined}
 as ``BP part'' and to the factor graph representing the factorization $\prod_{a\in \A_{\text{MF}}}f_a(\xva)$ in \eqref{eq:factorcombined} as ``MF part''. Furthermore, we set
\begin{align*}
 \I_{\text{MF}}\triangleq\bigcup_{a\in\A_{\text{MF}}} \N(a),&&
 \I_{\text{BP}}\triangleq\bigcup_{a\in\A_{\text{BP}}} \N(a)
\end{align*}
and
\begin{align*}
\M_{\text{MF}}(i)\triangleq\A_{\text{MF}}\cap \M(i),&&
\M_{\text{BP}}(i)\triangleq\A_{\text{BP}}\cap \M(i).
\end{align*}

Next, we define the following regions and counting numbers:
\begin{enumerate}
\item one MF region $R_{\text{MF}}\triangleq(\I_{\text{MF}},\A_{\text{MF}})$, with  $c_{R_{\text{MF}}}=1$;
\item small regions $R_i\triangleq(\{i\},\emptyset)$, with $c_{R_i}=1-|\M_{\text{BP}}(i)|-
\IND_{\I_\text{MF}}(i)$
for all $i\in\I_{\text{BP}}$;
\item large regions $R_a\triangleq(\N(a),\{a\})$, with $c_{R_a}=1$ for all $a\in\A_{\text{BP}}$.
\end{enumerate}
This yields the valid set of regions and associated counting numbers
\begin{align}\label{Rcombined}
\R_\text{BP,\,MF}
\triangleq&\, \{(R_i,c_{R_i})\mid i\in\I_{\text{BP}}\}\cup\{(R_a,c_{R_a})\mid a\in\A_\text{BP}\}\nonumber\\
&\, \cup\{(R_\text{MF}, c_{R_\text{MF}})\}.
\end{align}
The additional terms $\IND_{\I_\text{MF}}(i)$ in the counting numbers of the small regions $R_i$ $(i\in\I)$  defined in 2) compared to the counting numbers of the small regions for the Bethe approximation
(see Example \ref{example2}) guarantee that $\R_\text{BP,\,MF}$ is indeed a valid set of regions and associated counting numbers.

The valid set of regions and associated counting numbers in \eqref{Rcombined} gives the region-based free energy approximation
\begin{align}\label{Fcombined}
\! F_\text{BP,\,MF}=&
\sum_{a\in\A_{\text{BP}}}\sum_{\xva}b_a(\xva)\ln\frac{b_a(\xva)}{f_a(\xva)}\nonumber\\
&-\sum_{a\in\A_{\text{MF}}}\sum_{\xva}\prod_{i\in \N(a)} b_i(x_i)\ln f_a(\xva)\nonumber\\
&-\sum_{i\in\I}(|\M_{\text{BP}}(i)|-1)\sum_{x_i}b_i(x_i)\ln b_i(x_i)
\end{align}
with $F_\text{BP,\,MF}\triangleq F_{\R_\text{BP,\,MF}}$.
In \eqref{Fcombined}, we have already plugged in the factorization constraint 
\begin{equation*}
b_\text{MF}(\xv_\text{MF})=\prod_{i\in\I_\text{MF}}b_i(x_i)
\end{equation*}
with $\xv_\text{MF}\triangleq(x_i\mid i\in \I_{\text{MF}})^{\operatorname{T}}$ and $b_\text{MF}\triangleq b_{R_\text{MF}}$.
The beliefs $b_i$ $(i\in\I)$ and $b_a$ $(a\in\A_\text{BP})$ have to fulfill the normalization constraints 
\begin{align}\label{eq:normalizationconstrains2}
\begin{split}
\sum_{x_i}b_i(x_i)&=1,\quad\text{for all}\ i\in\I_{\text{MF}}\setminus\I_{\text{BP}}\\
\sum_{\xva} b_a(\xva)&=1,\quad\text{for all}\ a\in\A_{\text{BP}}
\end{split}
\end{align}
and the marginalization constraints
\begin{align}\label{eq:marginalizationconstrains2}
b_i(x_i)=\sum_{\xva\setminus x_i}b_a(\xva),\quad\text{for all}\ a\in\A_{\text{BP}}, i\in\N(a).
\end{align}
\begin{remark}
Note that there is no need to introduce normalization constraints for the beliefs $b_i$ $(i\in\I_\text{BP})$.  If $a\in\M_\text{BP}(i)$, then
it follows from the normalization constraint for the belief $b_a$ and marginalization constraint for the beliefs $b_a$ and $b_i$ that
\begin{align*}
1&=\sum_{\xva} b_a(\xva)\\
&=\sum_{x_i}\Big(\sum_{\xva\setminus x_i} b_a(\xva)\Big)\\
&=\sum_{x_i} b_i(x_i).
\end{align*}
\end{remark}
We will show in Lemma \ref{theohard} that the region-based free energy approximation in  \eqref{Fcombined} fulfilling the constraints  \eqref{eq:normalizationconstrains2} and \eqref{eq:marginalizationconstrains2} is a finite quantity, i.e., that $-\infty < F_\text{BP,\,MF} < \infty$.

The constraints \eqref{eq:normalizationconstrains2} and \eqref{eq:marginalizationconstrains2} can  be  included in the Lagrangian  \cite[Sec.\,3.1.3]{be03}
\begin{align}\label{Lcombined}
 L_\text{BP,\,MF}\triangleq&\, F_\text{BP,\,MF}\nonumber\\
 &-\sum_{a\in\A_{\text{BP}}}\sum_{i\in \N(a)}\sum_{x_i}\lambda_{a,i}(x_i)
\Big(b_i(x_i)-\sum_{\xva\setminus x_i}b_a(\xva)\Big)\nonumber\\
&-\sum_{i\in\I_\text{MF}\setminus\I_\text{BP}}\gamma_i\Big( \sum_{x_i}b_i(x_i) -1\Big)\nonumber\\
&-\sum_{a\in\A_\text{BP}}\gamma_a\Big( \sum_{\xva}b_a(\xva) -1\Big)
.
\end{align}
The stationary points of the Lagrangian $L_\text{BP,\,MF}$  in \eqref{Lcombined} are then obtained by setting the derivatives of
$L_\text{BP,\,MF}$ with respect to the beliefs and the Lagrange multipliers equal to zero. The following theorem relates the stationary points of the Lagrangian $L_\text{BP,\,MF}$ to solutions of fixed-point equations for the beliefs.

\begin{theorem}\label{Theoremcombined}
Stationary points of the Lagrangian in \eqref{Lcombined} in the combined BP--MF approach must be
fixed-points with positive beliefs fulfilling
\begin{equation}
\label{bcombined}
\begin{cases}
\begin{split}
b_a(\xva)=&\, \d_a\,
f_a(\xva)\prod\limits_{i\in \N(a)}\n_{i\to a}(x_i),\\
&\, \text{for all}\ a\in\A_\text{BP}\\
b_i(x_i)=&\, \d_i\!\!\!\!
\prod\limits_{a\in \M_{\text{BP}}(i)}\m^{\text{BP}}_{a\to i}(x_i)
\prod\limits_{a\in \M_{\text{MF}}(i)}\m^{\text{MF}}_{a\to i}(x_i),\\
&\, \text{for all}\ i\in\I
\end{split}
\end{cases}
\end{equation}
with
\begin{equation}
\label{FPcombined}
\begin{cases}
\begin{split}
\n_{i\to a}(x_i)=&\, \d_i
\prod\limits_{c\in \M_{\text{BP}}(i)\setminus a}\m^{\text{BP}}_{c\to i}(x_i)
\prod\limits_{c\in \M_{\text{MF}}(i)}m^{\text{MF}}_{c\to i}(x_i),\\
&\, \text{for all}\ a\in\A, i\in\N(a)\\
\m^{\text{BP}}_{a\to i}(x_i)=&\d_a
\sum_{\xva\setminus x_i} f_a(\xva) \prod_{j\in \N(a)\setminus i} \n_{j\to a}(x_j),\\
&\, \text{for all}\  a\in \A_\text{BP}, i\in\N(a) \\
m^{\text{MF}}_{a\to i}(x_i)=&
\exp\Biggl(\ \sum_{\xva\setminus x_i}\prod_{j\in \N(a)\setminus i}\n_{j\to a}(x_j)\ln f_a(\xva)\Biggr),\\
&\, \text{for all}\  a\in \A_\text{MF}, i\in\N(a)
\end{split}
\end{cases}
\end{equation}
and vice versa. Here, $\d_i$ $(i\in\I)$ and  $\d_a$ $(a\in\A_\text{BP})$ are positive constants that ensure that the beliefs $b_i$ $(i\in\I)$  and $b_a$ $(a\in\A)$
are normalized to one with  $\d_i=1$ for all $i\in\I_\text{BP}$.
\end{theorem}
\proof
See Appendix \ref{Proofcombined}.\endproof

\begin{remark}\label{pdf}
Note that for each $k\in\I\setminus\I_\text{BP}$ Theorem \ref{Theoremcombined} can be generalized to the case where $X_k$ is a continuous random variable following the
derivation presented in Appendix \ref{AppcontMF}. Formally, each sum over $x_k$ with $k\in\I\setminus\I_\text{BP}$ in the third identity in  \eqref{FPcombined} has to be replaced by a Lebesgue integral whenever the corresponding random variable $X_k$ is continuous.
\end{remark}

\begin{remark}\label{remarkextapp}
Note that Theorem \ref{Theoremcombined} clearly states whether ``extrinsic"
values or ``APPs"  should be passed. 
In fact, the first equation in \eqref{FPcombined} implies that each message $n_{i\to a}(x_i)$ $(a\in\A, i\in\I)$ is an ``extrinsic" value when $a\in\A_\text{BP}$ and an ``APP" when $a\in\A_\text{MF}$.
\end{remark}

\subsection{Hard constraints for BP}\label{hard}
Some suggestions on how to generalize Theorem \ref{the1} (\!\!\!\cite[Th.\,2]{yefrwe04}) to hard constraints, i.e., to the case where the factors
of the pmf $p_{\Xm}$ are not restricted to be strictly positive real-valued functions, can be found in \cite[Sec.\,VI.D]{yefrwe04}. An example of hard constraints are deterministic functions like, e.g., code constraints. However, the statements formulated there are only conjectures and are based on the assumption that we can always compute the derivative of the Lagrange function with respect to the beliefs. This is not always possible
because
\[
\frac{\partial F_\text{BP}}{\partial b_a(\xva)} \to \infty,\quad \text{as}\ f_a(\xva)\to 0
\]
with $F_\text{BP}$ from \eqref{bethefree}.
In the sequel, we show how to  generalize Theorem \ref{Theoremcombined} to the case where $f_a\geq 0$ for all $a\in\A_{\text{BP}}$ based on the simple observation that we are interested in solutions where the region-based free energy approximation is not plus infinity (recall that we want to minimize this quantity). As a byproduct, this also yields an extension of Theorem \ref{the1} (\!\!\cite[Th.\,2]{yefrwe04}) to hard constraints by simply setting $\A_\text{MF}=\emptyset$.

\begin{lemma}\label{theohard}
Suppose that
\begin{align}
f_a&\geq 0,\quad\text{for all}\  a\in\A_\text{BP}\label{eq:faBP}\\
f_a&> 0,\quad\text{for all}\  a\in\A_\text{MF}\label{eq:faMF}
\end{align}
and $p_{\Xm}\mid_{\bar x_i}\not \equiv 0$ for all $i\in\I$ and  each realization $\bar x_i$ of $X_i$.\footnote{If $p_{\Xm}\mid_{\bar x_i} \equiv 0$ then we can simply remove this realization $\bar x_i$ of $X_i$.}
Furthermore, we assume  that $b_i$ $(i\in\I)$ and $b_a$ $(a\in\A_\text{BP})$ fulfill the constraints \eqref{eq:normalizationconstrains2} and \eqref{eq:marginalizationconstrains2}.
Then
\begin{enumerate}
\item $F_\text{BP,MF}>-\infty$;
\item The condition
\begin{align}
b_a(\xvba)&=0, \quad\text{for all}\ \xvba\ \text{with}\  a\in\A_{\text{BP}}, f_a(\xvba)=0\label{eq:removestate1}
\end{align}
is  necessary and sufficient for $F_\text{BP,MF}<\infty$;
\item If  \eqref{eq:removestate1} is fulfilled, the remaining stationary points $b_i(x_i)$ ($i\in\I$) and $b_a(\xva)$ excluding all
$\xvba$ from \eqref{eq:removestate1} ($a\in\A_\text{BP}$)  of the Lagrangian in \eqref{Lcombined}
are positive beliefs fulfilling \eqref{bcombined} and \eqref{FPcombined} excluding all
$\xvba$ from \eqref{eq:removestate1} and vice versa.
\item
Moreover, \eqref{bcombined} and \eqref{FPcombined} hold for all realizations $\xvba$  (including all
$\xvba$ from \eqref{eq:removestate1}) and, therefore,
\eqref{bcombined} contains \eqref{eq:removestate1} as a special case.
\end{enumerate}
\end{lemma}
\proof
See Appendix \ref{proofhard}.\endproof

\begin{remark}
At first sight it seems to be a contradiction to the marginalization constraints  \eqref{eq:marginalizationconstrains2} that \eqref{eq:removestate1}  holds and all the beliefs $b_i$ ($i\in\I_\text{BP}$) are strictly positive functions. To illustrate that this is indeed the case, let $i\in\I_\text{BP}$, $a\in\M_\text{BP}(i)$, and fix one realization $\bar x_i$ of $X_i$. Since $p_{\Xm}\mid_{\bar x_i}\not\equiv 0$ we also have $f_a\mid_{\bar x_i}\not\equiv 0$. This implies that $f_a(\xvba)\not=0$ for at least one realization $\xvba=(\bar x_j\mid j\in\N(a))^{\operatorname{T}}$ with $i\in\N(a)$ and, therefore,  $b_a(\xvba)\neq 0$.  The marginalization constraints \eqref{eq:marginalizationconstrains2} together with the fact that the belief $b_a$ must be a nonnegative function then implies that we have indeed $b_i(\bar x_i)>0$.
\end{remark}


\subsection{Convergence and main algorithm}\label{Rconvergence}

If the  BP part has no cycle 
and
\begin{align}\label{eq:condconv}
|\N(a)\cap\I_\text{BP}|&\leq 1,\quad\text{for all}\ a\in\A_\text{MF}
\end{align}
then there exists a convergent implementation of the combined message passing equations in \eqref{FPcombined}. In fact, we can iterate between
updating the beliefs $b_i$ with $i\in\I_{\text{MF}}\setminus\I_{\text{BP}}$ and the forward backward algorithm in the BP part, as  outlined in the following Algorithm.
\begin{algorithm}\label{Algorithm}
If the BP part has no cycle and
\eqref{eq:condconv} is fulfilled, the following implementation of the fixed-point equations in \eqref{FPcombined} is guaranteed to converge.
\begin{enumerate}
\item Initialize $b_i$ for all $i\in\I_{\text{MF}}\setminus\I_\text{BP}$ and send the corresponding messages
$n_{i\to a}(x_i)=b_i(x_i)$ to all factor nodes $a\in\M_{\text{MF}}(i)$.
\item
Use all messages
$m^{\text{MF}}_{a\to i}(x_i)$ with $i\in\I_{\text{BP}}\cap\I_{\text{MF}} $ and $a\in\M_{\text{MF}}(i)$ as fixed input for the BP part and
run the forward/backward algorithm \cite{ksbrlo01}. The fact that the resulting beliefs $b_i$ with $i\in\I_{\text{BP}}$ cannot increase the region-based free energy approximation in \eqref{Fcombined} is proved in Appendix \ref{Aconvergence}.
\item For each $i\in \I_{\text{MF}}\cap\I_{\text{BP}}$ and  $a\in\M_{\text{MF}}(i)$ the message $n_{i\to a}(x_i)$ is now available and can
be used for further updates in the MF part.
\item
For each $i\in\I_{\text{MF}}\setminus\I_{\text{BP}}$   successively recompute the message $n_{i\to a}(x_i)$ and send it to all $a\in\M_{\text{MF}}(i)$.
Note that for all indices $i\in\I_{\text{MF}}\setminus\I_{\text{BP}}$ 
\begin{align*}
\frac{\partial^2 F_\text{BP,\,MF}}{\partial b_i(x_i)^2} =\frac{1}{b_i(x_i)}>0
\end{align*}
and  the set of all beliefs $b_i$ satisfying the normalization constraint (first equation in \eqref{eq:normalizationconstrains2}) 
 is a convex set. This implies that for each $i\in\I_{\text{MF}}\setminus\I_{\text{BP}}$ we are solving a convex optimization problem. 
 Therefore, the region-based free energy approximation in \eqref{Fcombined} cannot increase.
\item Proceed as described in 2).
\end{enumerate}
\end{algorithm}
\begin{remark}\label{RemarkloopyBP}
If the factor graph representing the BP part is not cycle-free then Algorithm \ref{Algorithm} can be modified by running loopy BP in step 2). However, in this case the algorithm is not guaranteed to converge.
\end{remark}

\section{Application to iterative channel estimation and decoding}\label{Application}
In this section, we present an example where we show how to compute the updates of the messages in \eqref{FPcombined} based on Algorithm \ref{Algorithm}. We choose a simple communication model where the updates of the messages are simple enough in order to avoid overstressed notation. A class of more complex MIMO-OFDM receiver architectures together with numerical simulations can be found in  \cite{makirichfl11}. In our example, we use BP for modulation and decoding and the MF approximation for estimating the parameters of the a posteriori distribution of the channel gains. This splitting is convenient because BP works well with hard constraints and the MF approximation yields very simple message passing update equations due to the fact that the MF part in our example is a conjugate-exponential model \cite{CWi05}. Applying BP to all factor nodes  would be intractable because the complexity is too high, cf. the discussion in Subsection \ref{secBPE}.

Specifically, we consider an OFDM system with $M+N$ active subcarriers.
We denote by $\D\subset{[1:M+N]}$ and $\P\subset[1:M+N]$ the sets of subcarrier indices for the data and pilot symbols, respectively with $|\P|=M$, $|\D|=N$, and $\P\cap\D=\emptyset$. 

In the transmitter, a random vector $\Um=(U_i\mid i\in [1:K])$ representing the information bits is encoded and interleaved using a rate $R=K/(LN)$ encoder and a random interleaver, respectively into the random vector 
\[
\Cm={\Big({\Cm^{(1)}}^{\operatorname{T}},\dots,{\Cm^{(N)}}^{\operatorname{T}}\Big)}^{\operatorname{T}}
\]
of length $LN$  representing the coded and interleaved bits. Each random subvector 
$\Cm^{(n)}\triangleq (C_1^{(n)},\dots,C_L^{(n)})^{\operatorname{T}}$  of length $L$ is then  mapped, i.e., modulated,  
to $X_{i_n}\in\mathcal{S}$ with $i_n\in\D$  $(n\in [1:N])$, where $\mathcal{S}$ is a complex modulation alphabet of size $|\mathcal{S}|=2^L$. 

After removing the cyclic prefix in the receiver, we get the following input-output relationship in the frequency domain:
\begin{equation}\label{eq:channel}
\begin{split}
\Ym_\text{D}&=\Hm_\text{D}\odot\Xm_\text{D}+\Zm_\text{D}\\
\Ym_\text{P}&=\Hm_\text{P}\odot\xv_\text{P}+\Zm_\text{P}
\end{split}
\end{equation}
where 
$\Xm_\text{D}\triangleq(X_i\mid i\in \D)^{\operatorname{T}}$ is the random vector corresponding to the transmitted data symbols, 
$\xv_\text{P}\triangleq(x_i\mid i\in \P)^{\operatorname{T}}$ is the vector containing the transmitted pilot symbols, and 
$\Hm_\text{D}\triangleq(H_i\mid i\in \D)^{\operatorname{T}}$ and $\Hm_\text{P}\triangleq(H_i\mid i\in \P)^{\operatorname{T}}$ are 
random vectors representing the multiplicative action of the channel 
while $\Zm_\text{D}\triangleq(Z_i\mid i\in \D)^{\operatorname{T}}$ and $\Zm_\text{P}\triangleq(Z_i\mid i\in \P)^{\operatorname{T}}$ are 
random vectors representing additive Gaussian noise with $p_{\mathbf{Z}}(\zv)=\text{CN}(\mathbf{z};\mathbf{0},\gamma^{-1}\mathbf{I}_{M+N})$ and   
 $\Zm\triangleq(Z_i\mid i\in \D\cup\P)^{\operatorname{T}}$.
Note that \eqref{eq:channel} is very general and can also be used to model, e.g., a time-varying frequency-flat channel. 

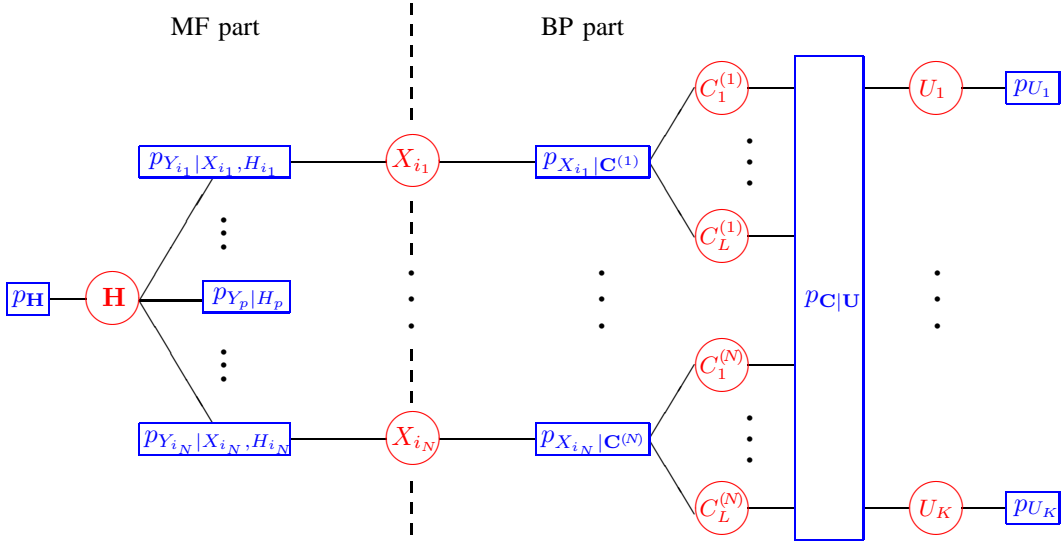
\begin{figure*}[htbp]
\begin{center}
\begin{picture}(300,190)(100,-35)
\color{black}
\put(110,150){MF part}
\put(250,150){BP part}
\put(98,49.5){\line(3,5){27.8}}
\put(98,49.5){\line(3,-5){27.8}}
\put(155,-3){\line(1,0){36.5}}
\put(155,102){\line(1,0){36.5}}
\put(211.5,-3){\line(1,0){36.5}}
\put(211.5,102){\line(1,0){36.5}}
\put(291,-3){\line(3,5){17}}
\put(291,-3){\line(3,-5){17}}
\put(291,102){\line(3,5){17}}
\put(291,102){\line(3,-5){17}}
\put(328,130){\line(1,0){18}}
\put(328,74){\line(1,0){18}}
\put(328,25){\line(1,0){18}}
\put(328,-29){\line(1,0){18}}
\put(64,50){\line(1,0){14}}
\put(372,130){\line(1,0){17}}
\put(372,-29){\line(1,0){17}}
\put(409,130){\line(1,0){17}}
\put(409,-29){\line(1,0){17}}
\put(98,49.5){\line(1,0){24}}
\put(201,158){\multiput(0,0)(0,-8){6}{\line(0,1){4}}}
\put(201,83){\multiput(0,0)(0,-8){3}{\line(0,1){4}}}
\put(201,27){\multiput(0,0)(0,-8){3}{\line(0,1){4}}}
\put(201,-23){\multiput(0,0)(0,-8){3}{\line(0,1){4}}}

\put(130,75)
{
\put(0,5){\circle*{2}}
\put(0,0){\circle*{2}}
\put(0,-5){\circle*{2}}
}

\put(130,25)
{
\put(0,5){\circle*{2}}
\put(0,0){\circle*{2}}
\put(0,-5){\circle*{2}}
}

\put(201,50)
{
\put(0,10){\circle*{2}}
\put(0,0){\circle*{2}}
\put(0,-10){\circle*{2}}
}
\put(273,50)
{
\put(0,10){\circle*{2}}
\put(0,0){\circle*{2}}
\put(0,-10){\circle*{2}}
}
\put(329,102)
{
\put(0,8){\circle*{2}}
\put(0,0){\circle*{2}}
\put(0,-8){\circle*{2}}
}
\put(329,-3)
{
\put(0,8){\circle*{2}}
\put(0,0){\circle*{2}}
\put(0,-8){\circle*{2}}
}
\put(400,50)
{
\put(0,10){\circle*{2}}
\put(0,0){\circle*{2}}
\put(0,-10){\circle*{2}}
}
\color{blue}
\put(50,49)
{
\put(-2,-5){\line(1,0){16}}
\put(-2,7){\line(1,0){16}}
\put(-2,-5){\line(0,1){12}}
\put(14,-5){\line(0,1){12}}
$p_\Hm$
}
\put(100,101)
{
\put(-2,-5){\line(1,0){57}}
\put(-2,7){\line(1,0){57}}
\put(-2,-5){\line(0,1){12}}
\put(55,-5){\line(0,1){12}}
\put(2,0){$p_{Y_{i_1}\mid X_{i_1},H_{i_1}}$}
}
\put(100,-4)
{
\put(-2,-5){\line(1,0){57}}
\put(-2,7){\line(1,0){57}}
\put(-2,-5){\line(0,1){12}}
\put(55,-5){\line(0,1){12}}
$p_{Y_{i_N}\mid X_{i_N},H_{i_N}}$
}
\put(124,50)
{
\put(-2,-5){\line(1,0){33}}
\put(-2,7){\line(1,0){33}}
\put(-2,-5){\line(0,1){12}}
\put(31,-5){\line(0,1){12}}
\put(2,0){$p_{Y_{p}\mid H_p}$}
}

\put(252,101)
{
\put(-4,-5){\line(1,0){43}}
\put(-4,7){\line(1,0){43}}
\put(-4,-5){\line(0,1){12}}
\put(39,-5){\line(0,1){12}}
\put(-1,0){$p_{X_{i_1}\mid \Km^{(1)}}$}
}
\put(252,-4)
{
\put(-4,-5){\line(1,0){43}}
\put(-4,7){\line(1,0){43}}
\put(-4,-5){\line(0,1){12}}
\put(39,-5){\line(0,1){12}}
\put(-2,0){$p_{X_{i_N}\mid \Km^{(\!N\!)}}$}
}

\put(350,49)
{
\put(-4,-90){\line(0,1){183}}
\put(22,-90){\line(0,1){183}}
\put(-4,-90){\line(1,0){26}}
\put(-4,93){\line(1,0){26}}
$p_{\Km\mid \Um}$
}
\put(430,129)
{
\put(-4,-5){\line(1,0){20}}
\put(-4,7){\line(1,0){20}}
\put(-4,-5){\line(0,1){12}}
\put(16,-5){\line(0,1){12}}
\put(-1,0){$p_{U_{1}}$}
}
\put(430,-30)
{
\put(-4,-5){\line(1,0){20}}
\put(-4,7){\line(1,0){20}}
\put(-4,-5){\line(0,1){12}}
\put(16,-5){\line(0,1){12}}
\put(-1,0){$p_{U_{K}}$}
}

\color{red}
\put(85,48.5){\put(3,2.0){\circle{19}}
\put(-1,-1){$\Hm$}
}
\put(194.5,100){\put(7,2.5){\circle{19}}\put(-1,0){$X_{i_1}$}}
\put(194.5,-5){\put(7,2.5){\circle{19}}\put(-1,0){$X_{i_N}$}}
\put(311,128){
\put(7.5,2){\circle{20}}
\put(-1,-1){\small{$C^{(1)}_{1}$}}
}
\put(311,72){
\put(7.5,2){\circle{20}}
\put(-1,-1){\small{$C^{(1)}_{L}$}}
}
\put(311,23){
\put(7.5,2){\circle{20}}
\put(-1,-1){\small{$C^{(\!N\!)}_{1}$}}
}
\put(311,-31){
\put(7.5,2){\circle{20}}
\put(-1,-1){\small{$C^{(\!N\!)}_{L}$}}
}
\put(392,128){
\put(7.5,2){\circle{19}}
\put(1,-1){\small{$U_{1}$}}
}
\put(392,-31){
\put(7.5,2){\circle{19}}
\put(1,-1){\small{$U_{K}$}}
}

\end{picture}
\caption{Factor graph corresponding to the factorization of the pdf in \eqref{facexa} with $\D=\{i_1,\dots,i_N\}$ and $p\in\P$. The splitting of the factor graph into BP and MF part is chosen in such a way that utilizes most of the
 advantages of BP and the MF approximation. }
\label{Fig1}
\end{center}
\end{figure*}

Setting 
$\Ym\triangleq (Y_i\mid i\in\D\cup\P)^{\operatorname{T}}$ and 
$\Hm\triangleq (H_i\mid i\in\D\cup\P)^{\operatorname{T}}$,  
the pdf $p_{\Ym,\Xm_\text{D},\Hm,\Km,\Um}$ admits the factorization
\begin{align}\label{facexa}
&p_{\Ym,\Xm_\text{D},\Hm,\Km,\Um}(\yv,\xv_\text{D},\hv,\kv,\uv)\nonumber\\
&=p_{\Ym|\Xm_\text{D},\Hm}(\yv|\xv_\text{D},\hv)\,p_\Hm(\hv)\,p_{\Xm_\text{D}|\Km}(\xv_\text{D}|\kv)\,p_{\Km|\Um}(\kv|\uv)\,p_\Um(\uv)\nonumber\\
&=
\prod_{i\in\D}p_{Y_i|X_i,H_i}(y_i| x_i,h_i)\prod_{j\in\P}p_{Y_j|H_j}(y_j|h_j)\times p_\Hm(\hv)\nonumber\\
&\phantom{=}\times
\prod_{n\in[1:N]}p_{X_{i_n}|\Km^{(n)}}\big(x_{i_n}|\kv^{(n)}\big)\times p_{\Km|\Um}(\kv|\uv)\nonumber\\
&\phantom{=}\times \prod_{k\in [1:K]}p_{U_k}(u_k)
\end{align}
where we used the fact that $\Hm$ is independent of $\Xm_\text{D}$, $\Cm$, and $\Um$ and  $\Ym$ is independent of $\Cm$ and $\Um$ conditioned on $\Xm_\text{D}$.
Note that
\begin{align}
p_{Y_i|X_i,H_i}(y_i| x_i,h_i)
&=\frac{\gamma}{\pi}\exp(-\gamma |y_i-h_i x_i|^2)\nonumber\\
&=\text{CN}(y_i;h_i x_i,1/\gamma),\quad\text{for all}\ i\in \D\label{eq:pycond}\\
p_{Y_i|H_i}(y_i|h_i)
&=\frac{\gamma}{\pi}\exp(-\gamma |y_i-h_i x_i|^2)\nonumber\\
&=\text{CN}(y_i;h_i x_i,1/\gamma),\quad\text{for all}\  i\in \P.\label{eq:pycondP}
\end{align}
We choose for the prior distribution of $\Hm$ 
\begin{align*}
 p_\Hm(\hv)&=\text{CN}(\hv;\muv_\Hm^\text{P},{\Lam_\Hm^\text{P}}^{-1}).
\end{align*}
Now define
\begin{align}
\I\triangleq\,&
\{X_i\mid i\in\D\}\cup\{\Hm\}\nonumber\\
&\cup\{C^{(1)}_1,\dots,C^{(N)}_{L}\}\cup\{U_1,\dots,U_K\}
\label{eq:Iexample}\\
\A\triangleq\,&
\{p_{Y_i\mid X_i,H_i}\mid i\in \D\}
\cup
\{p_{Y_i\mid H_i}\mid i\in \P\}
\cup\{p_\Hm\}
\nonumber\\
&\cup \{p_{X_{i_n}\mid \Km^{(n)}}\mid n\in [1:N]\}\nonumber\\
&\cup \{p_{\Km|\Um}\}\cup\{p_{U_k}\mid k\in [1:K]\}
\label{eq:Aexample}
\end{align}
and set $f_a\triangleq a$ for all $a\in\A$. For example, we have $f_{p_\Hm}(\hv)=p_\Hm(\hv)$.
We choose a splitting of $\A$ into $\A_\text{BP}$ and $\A_\text{MF}$ with
\begin{equation}\label{eq:splittingexa}
\begin{split}
\A_\text{BP}\triangleq\,&
\{p_{X_{i_n}|\Km^{(n)}}\mid n\in [1:N]\}\\
&\cup\{p_{\Km|\Um}\}\cup\{p_{U_k}\mid k\in [1:K]\}\\
\A_\text{MF}\triangleq\,&
\{p_{Y_i\mid X_i,H_i}\mid i\in \D\}
\cup\{p_{Y_i\mid H_i}\mid i\in \P\}\cup\{p_\Hm\}.
\end{split}
\end{equation}
With this selection
\begin{equation*}
\begin{split}
\I_\text{BP}=&\,\{X_i\mid i\in\D\}\cup\{C^{(1)}_1,\dots,C^{(N)}_{L}\}\\
&\cup\{U_1,\dots,U_K\}\\
\I_\text{MF}=&\,\{X_i\mid i\in\D\}\cup\{\Hm\}
\end{split}
\end{equation*}
which implies that $\I_\text{BP}\cap\I_\text{MF}=\{X_i\mid i\in\D\}$. The factor graph corresponding to the factorization in \eqref{facexa} with the splitting of
$\A$ into $\A_\text{MF}$ and $\A_\text{BP}$ as in \eqref{eq:splittingexa} is depicted in Figure \ref{Fig1}.

We now show how to apply the variant of Algorithm \ref{Algorithm}  referred to in Remark \ref{RemarkloopyBP} to the factor graph depicted in Figure \ref{Fig1}. 
Note that \eqref{eq:condconv} is fulfilled in this example; however, cycles occur in the BP part of the factor graph due to the combination of 
(convolutional) coding, interleaving, and high-order modulation (see Table \ref{Parameters}).

\begin{algorithm} \label{Algorithm1}{\ } \\ 

\vspace*{-4truemm}
\begin{enumerate}
\item Initialize
\begin{align*}
b_{\Hm}(\hv)&=
\text{CN}(\hv;\muv_{\Hm},\Lam_{\Hm}^{-1})
\end{align*}
by setting 
\begin{align*}
\muv_{\Hm}&=\Lam_{\Hm}^{-1}(\Lam_{\Hm}^\text{P}\muv_{\Hm}^\text{P}+\widetilde\Lam_\Hm\widetilde\muv_\Hm)\\
\Lam_\Hm&=\Lam_{\Hm}^\text{P}+\widetilde\Lam_\Hm
\end{align*}

with 

\begin{align*}
\widetilde\lambda_{\Hm_{ij}}
&=
\begin{cases}
\gamma|x_i|^2 &\text{if}\ i=j\in\P\\
0&\text{else}
\end{cases}
\end{align*}
and
\begin{align*}
\widetilde\lambda_{\Hm_{ii}}\widetilde\mu_{\Hm_{i}}
&=
\begin{cases}
\gamma y_i x_i^\ast&\text{if}\ i\in\P\\
0&\text{if}\ i\in\D
\end{cases}
\end{align*}

and set 
\begin{align*}
n_{\Hm\to p_{Y_i\mid X_i,H_i}}(\hv)&=b_{\Hm}(\hv),\quad\text{for all}\  i\in \D.
\end{align*}
\item
Using the particular form of the distributions
$p_{Y_i\mid X_i,H_i}$ $(i\in \D)$ in \eqref{eq:pycond} and
$p_{Y_i\mid H_i}$ $(i\in \P)$ in \eqref{eq:pycondP},
compute 
\begin{align*}
&m^\text{MF}_{p_{Y_i\mid X_i,H_i}\to X_i}(x_i)\\
&\propto\exp\Bigg(-\gamma\int \operatorname{d}\!\hv\,n_{\Hm\to p_{Y_i\mid X_i,H_i}}(\hv)|y_i-h_ix_i|^2\Bigg)\\
&\propto\exp\Bigg(-\gamma(\sigma_{H_i}^2+|\mu_{H_i}|^2)\Bigg|x_i- \frac{y_i\mu_{H_i}^\ast}{\sigma_{H_i}^2+|\mu_{H_i}|^2}\Bigg|^2\Bigg)\\
&\propto\text{CN}\Bigg(x_i;\frac{y_i\mu_{H_i}^\ast}{\sigma_{H_i}^2+|\mu_{H_i}|^2},\frac{1}{\gamma(\sigma_{H_i}^2+|\mu_{H_i}|^2)}\Bigg)
\end{align*}
for all $i\in\D$ with $\sigma^2_{H_i}\triangleq[\Lam_{\Hm}^{-1}]_{i,i}$ $(i\in \D)$.

\item Use the messages $m^\text{MF}_{p_{Y_i\mid X_i,H_i}\to X_i}(x_i)$ $(i\in \D)$ as fixed input for the BP part and run BP.
\item After running  BP in the BP part, compute the messages $n_{X_{i}\to p_{Y_{i}\mid X_{i},H_i}}(x_i)$ $(i\in\D)$ and update the messages in the MF part.
Namely, after setting
\begin{align*}
\mu_{X_{i}}&\triangleq \sum_{x_{i}}n_{X_{i}\to p_{Y_{i}\mid X_{i},H_i}}(x_{i}) x_{i}\\
\sigma^2_{X_{i}}&\triangleq \sum_{x_{i}}n_{X_{i}\to p_{Y_{i}\mid X_{i},H_i}}(x_{i}) |x_{i}-\mu_{X_{i}}|^2
\end{align*}
for all $i\in\D$, compute the messages
\begin{align*}
&m^{\text{MF}}_{p_{Y_i\mid X_i,H_i}\to \Hm}(h_i)\\
&\propto
\exp
\Bigg(
-\gamma\sum_{x_i}
n_{X_i\to p_{Y_i\mid X_i,H_i}}(x_i)
|y_i-h_ix_i|^2
\Bigg)\nonumber\\
&\propto
\exp\Bigg(-\gamma(\sigma_{X_i}^2+|\mu_{X_i}|^2)\Bigg|h_i-\frac{y_i\mu_{X_i}^\ast}{\sigma_{X_i}^2+|\mu_{X_i}|^2} \Bigg|^2\Bigg)\\
&\propto\text{CN}\Bigg(h_i;\frac{y_i\mu_{X_i}^\ast}{\sigma_{X_i}^2+|\mu_{X_i}|^2},\frac{1}{\gamma(\sigma_{X_i}^2+|\mu_{X_i}|^2)}\Bigg)
\end{align*}
for all  $i\in\D$, 
\begin{align*}
m^{\text{MF}}_{p_{Y_i\mid H_i}\to \Hm}(h_i)
&\propto \exp(-\gamma |y_i-h_ix_i|^2)\\
&\propto\text{CN}\Bigg(h_i;\frac{y_ix_i^\ast}{|x_i|^2},\frac{1}{\gamma |x_i|^2)}\Bigg)  
\end{align*}
for all $i\in\P$,
\begin{align*}
m^{\text{MF}}_{p_{\Hm}\to \Hm}(\hv)=\text{CN}(\hv;\muv_\Hm^\text{P},{\Lam_\Hm^\text{P}}^{-1})
\end{align*}
and
\begin{align*}
&n_{\Hm\to p_{Y_i\mid X_i,H_i}}(\hv)\\
&=\e_{\Hm}
\prod_{i\in\D}m^{\text{MF}}_{p_{Y_i\mid X_i,H_i}\to \Hm}(h_i)
\prod_{j\in\P}m^{\text{MF}}_{p_{Y_i\mid H_i}\to \Hm}(h_j)\\
&\phantom{=}\ \times m^{\text{MF}}_{p_{\Hm}\to \Hm}(\hv)\\
&=\frac{\det(\Lam_\Hm)}{\pi^{M+N}}\exp\Big(-(\hv-\mu_{\Hm})^{\operatorname{H}}\Lam_\Hm(\hv-\mu_{\Hm})\Big)\\
&=\text{CN}(\hv;\mu_{\Hm},\Lam^{-1}_\Hm)
\end{align*}
for all $i\in\D$.
Here, we used  Lemma \ref{lemmagaussian} in Appendix \ref{Agaussian} to get the updated parameters
\begin{equation}\label{eq:uH}
\begin{split}
\muv_{\Hm}&=\Lam_{\Hm}^{-1}(\Lam_{\Hm}^\text{P}\muv_{\Hm}^\text{P}+\widetilde\Lam_\Hm\widetilde\muv_\Hm)\\
\Lam_\Hm&=\Lam_{\Hm}^\text{P}+\widetilde\Lam_\Hm
\end{split}
\end{equation}
with
\begin{align*}
\widetilde\lambda_{\Hm_{ij}}
&=
\begin{cases}
\gamma(\sigma_{X_i}^2+|\mu_{X_i}|^2)&\text{if}\ i=j\in\D\\
\gamma|x_i|^2 &\text{if}\ i=j\in\P\\
0&\text{else}
\end{cases}
\end{align*}
and
\begin{align*}
\widetilde\lambda_{\Hm_{ii}}\widetilde\mu_{\Hm_{i}}
&=
\begin{cases}
\gamma y_i\mu_{X_i}^\ast&\text{if}\ i\in\D\\
\gamma y_i x_i^\ast&\text{if}\ i\in\P.
\end{cases}
\end{align*}

The update for the belief $b_{\Hm}$ is
\begin{align*}
b_{\Hm}(\hv)
&=n_{\Hm\to p_{Y_i\mid X_i,H_i}}(\hv)
\end{align*}
i.e., $b_{\Hm}(\hv)=\text{CN}(\hv;\mu_{\Hm},\Lam_{\Hm}^{-1})$.
\item Proceed as described in 2).
\end{enumerate}
\end{algorithm}

\subsection{``Extrinsic" values versus ``APP"}
In consideration of Remark \ref{remarkextapp} it is instructive to analyze the messages coming from the variable nodes
$\I_\text{BP}\cap\I_\text{MF}=\{X_1,\dots,X_{N}\}$, which are contained in the BP and MF part of the factor graph depicted in Figure \ref{Fig1}.
Whether a message passing from a variable node to a factor node is an ``extrinsic" value or an ``APP"
depends on whether the corresponding factor node is in the BP or the MF part. Thus, for all $n\in[1:N]$, the messages
\begin{align*}
n_{X_{i_n}\to p_{X_{i_n}\mid \Cm^{(n)}}}(x_{i_n})
=m^\text{MF}_{p_{Y_{i_n}\mid X_{i_n}, H_{i_n}}\to X_{i_n}}(x_{i_n})
\end{align*} 
which are passed into the BP part, are ``extrinsic" values, whereas the messages
\begin{align*}
&n_{X_{i_n}\to p_{Y_{i_n}\mid X_{i_n},H_{i_n}}}(x_{i_n})\\
&=m^\text{BP}_{p_{X_{i_n}\mid \Cm^{(n)}}\to X_{i_n}}(x_{i_n})\ m^\text{MF}_{p_{Y_{i_n}\mid X_{i_n},H_{i_n}}\to X_{i_n}}(x_{i_n})
\end{align*}
which are passed into the MF part, are ``APPs". Note that this result is aligned with the strategies proposed in \cite{weme06,romu08}, 
where ``APPs" are used for channel estimation and ``extrinsic values" for detection.

\subsection{Level of MF approximation}
Note that there is an ambiguity in the choice of variable nodes in the MF part. This ambiguity reflects the ``level of the MF approximation" and
results in a family of different algorithms. For example, instead of choosing $\Hm$ as a single random variable, we could have
chosen $H_i$ $(i\in[1:M+N])$ to be separate variable nodes in the factor graph. In this case we make the assumption that the random variables
$H_i$ $(i\in[1:M+N])$ are independent  and the set of indices $\I$ in
\eqref{eq:Iexample} has to be replaced by
\begin{align*}
\I\triangleq\,&
\{X_i\mid i\in\D\}\cup\{H_i\mid i\in\D\cup\P\}\nonumber\\
&\cup\{C^{(1)}_1,\dots,C^{(N)}_{L}\}\cup\{U_1,\dots,U_K\}.
\end{align*}
Since this is an additional approximation, the performance of the receiver is expected to decrease compared to the case where we choose $\Hm$ as a single random variable. However, it is possible that the complexity reduces by applying an additional MF approximation. See \cite{makirichfl11} for further discussions on this ambiguity for a class of  MIMO-OFDM receivers.

\subsection{Comparison with BP combined with Gaussian approximation}\label{secBPE}
The example makes evident how the complexity of the message passing algorithm can be simplified by exploiting the conjugate-exponential property of the MF part, which leads to simple update equations of the belief  $b_{\Hm}$. In fact, at each iteration in the algorithm we only have to update the parameters of a Gaussian distribution \eqref{eq:uH}.
In comparison let us consider an alternative split of $\A$ by moving the factor nodes $p_{Y_i\mid X_i,H_i}$ $(i\in \D)$ in \eqref{eq:pycond} and $p_{Y_i\mid H_i}$ $(i\in \P)$ in \eqref{eq:pycondP} to the BP part. This is equivalent to applying BP to the whole factor graph in Figure \ref{Fig1} because $m^\text{MF}_{p_{\Hm}\to\Hm}=m^\text{BP}_{p_{\Hm}\to\Hm}$. 
Doing so, each message $m^{\text{BP}}_{p_{Y_i\mid X_i,H_i}\to \Hm}(h_i)$ $(i\in\D)$ does no longer admit a closed form expression in terms of the mean and the variance of the  random variable $X_i$ and 
becomes a mixture of Gaussian pdfs with $2^L$ components; in consequence, each message 
$n_{\Hm\to p_{Y_i\mid X_i,H_i}}(\hv)$ $(i\in \D)$ becomes a sum of $2^{L(N-1)}$ terms. To keep the complexity of computing these messages tractable one has to rely on additional approximations.

\begin{figure}[htbp]
\includegraphics[width=1.0\columnwidth]{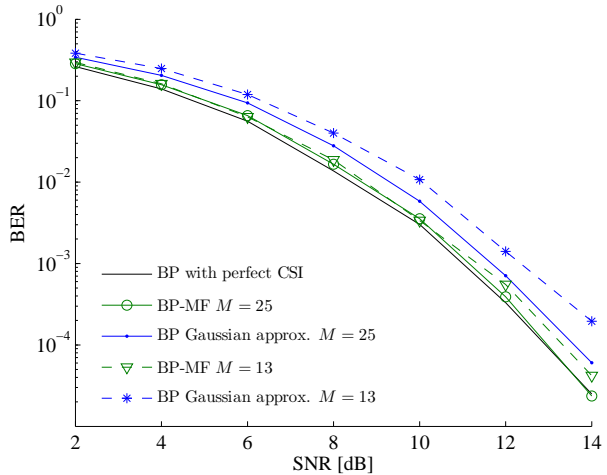}
\caption{Bit error rate (BER) as a function of signal-to-noise ratio (SNR) for Algorithm  \ref{Algorithm1} (BP--MF), BP combined with Gaussian approximation as described in Subsection \ref{secBPE}, and BP with perfect CSI at the receiver. 
Pilot spacing $\Delta_P\approx W_\text{coh}$ $(M=25)$ and $\Delta_P\approx 2W_\text{coh}$ $(M=13)$. 
}
\label{sim1}
\end{figure}

As suggested in \cite{knhotyau12,xu11}, we can approximate each message $m^{\text{BP}}_{p_{Y_i\mid X_i,H_i}\to \Hm}(h_i)$ $(i\in\D)$ by a Gaussian pdf. BP combined with this approximation is comparable in terms of complexity to Algorithm \ref{Algorithm1}, since the computations of the updates of the messages are equally complex. 
However, Algorithm  \ref{Algorithm1} clearly outperforms this alternative, as can be seen in Figure \ref{sim1}. 
It can also be noticed that the performance of Algorithm 2 is close to the case with perfect channel state information (CSI) at the receiver, even with a low density of pilots, i.e., such that the spacing  between any two consecutive pilots $(\Delta_P)$ approximately equals the coherence bandwidth\footnote{Calculated as the reciprocal of the maximum excess delay.} $(W_\text{coh})$ of the channel    or twice of it.

To circumvent the intractability of the BP-based receiver, one could also apply other approximate inference algorithms to the factor graph like, e.g., expectation propagation (EP). A comparison between EP and BP-MF can be found in \cite{bakimarifl12}, where it was shown that BP-MF yields the best performance-complexity tradeoff and does not suffer from numerical instability.

\begin{table}[htdp]
\caption{Parameters of the OFDM system.}
\begin{center}
\begin{tabular}{|l|l|}
\hline
Number of subcarriers &$M+N=300$\\
\hline
Number of evenly spaced pilots&$M\in\{13,25\}$\\
\hline
Modulation scheme for pilot symbols&$\text{QPSK}$\\
\hline
Modulation scheme for data symbols&$16\, \text{QAM}\ (L=4)$\\
\hline
Convolutional  channel code &$R=1/3$ $(133,171,165)_8$\\
\hline
Multipath channel model& $3\, \text{GPP}\ \text{ETU}$\\
\hline
Subcarrier spacing&$15\, \text{kHz}$\\
\hline
Coherence bandwidth & $W_\text{coh}\approx 200\,\text{kHz}$\\
\hline
\end{tabular}
\end{center}
\label{Parameters}
\end{table}%

\subsection{Estimation of noise precision}
Algorithm \ref{Algorithm1} can be easily extended to the case where the noise precision $\gamma$ is a realization of a random variable $\Gamma$. In fact, since $\ln p_{Y_i\mid X_i,H_i,\Gamma}$ $(i\in \D)$
and $\ln p_{Y_i\mid H_i,\Gamma}$ $(i\in \P)$ are linear in $\gamma$, we can replace any dependence on $\gamma$ in the existing messages in Algorithm \ref{Algorithm1} by the expected value of $\Gamma$ and get simple expressions for the additional messages  using a Gamma prior distribution for  $\Gamma$, reflecting the powerfulness of exploiting the conjugate-exponential model property in the MF part  for parameter estimation. See \cite{makirichfl11} for further details on the explicit form of the additional messages.

\section{Conclusion and Outlook}\label{Conclusion}
We showed that the message passing fixed-point equations of a combination of BP and the MF approximation correspond to stationary points of one single constrained region-based free energy approximation. These stationary points are in one-to-one correspondence to solutions of a coupled system of message passing fixed-point equations. For an arbitrary factor graph and a choice of a splitting of the factor nodes into a set of MF and BP factor nodes, our result gives immediately the corresponding message passing fixed-point equations and yields an interpretation of the computed beliefs as stationary points. Moreover, we presented an algorithm for updating the messages that is guaranteed to converge provided that the factor graph fulfills certain technical conditions. We also showed how to extend the MF part in the factor graph to continuous random variables and to include hard constraints in the BP part of the factor graph. 
Finally, we illustrated the computation of the messages of our algorithm in a simple example. This example demonstrates the efficiency of the combined scheme in models in which BP messages are computationally intractable. The proposed algorithm performs significantly better than the commonly used approach of using BP combined with a Gaussian approximation of computationally demanding messages.

An interesting extension of our result would be to generalize the BP part to contain also continuous random variables. The results in \cite{cr70} provide a promising approach. Indeed, they could be used to generalize the Lagrange multiplier for the marginalization constraints to the continuous case. However, these methods are based on the assumption that the objective function is Fr{\'e}chet differentiable \cite[p. 172]{lu97}. In general a region-base free energy approximation is neither Fr{\'e}chet differentiable nor Gateaux differentiable, at least not without any modification of the definitions used in standard text books
\cite[pp. 171--172]{lu97}\footnote{For a positive real-valued function $b$, $b+\Delta b$ might fail to be a positive real-valued function for arbitrary perturbations $\Delta b$ with sufficiently small norm $\|\Delta b\|$.}.
An extension to continuous random variables in the BP part would allow to apply a combination of BP with the MF approximation, e.g., for sensor self-localization, where both methods are used \cite{wyliwi97,pepefl11}. Another interesting extension could be to generalize the region-based free energy approximation such that the messages in the BP part are equivalent to the messages passed in tree reweighted BP or to include second order correction terms in the MF approximation that are similar to the Onsager reaction term \cite{ta00}.

\section{Acknowledgment}
The authors wish to thank Prof. Ralf R. M\"uller for his comments on a previous draft of this paper.

\appendix
\subsection{Proof of Lemma \ref{Lemmabetter}}\label{ProofLemmabetter}
Suppose that $\{\mt_{a\to i}(x_i), \nt_{i\to a}(x_i)\}$ $(a\in\A, i\in\N(a))$  is a solution of \eqref{eq:fp2a} and set
\begin{equation}\label{eq:rescaledmessages}
\begin{split}
\mt_{a\to i}(x_i)&=\kappa_{a, i}\m_{a\to i}(x_i),\quad\text{for all}\  a\in\A, i\in\N(a)\\
\nt_{i\to a}(x_i)&=\tau_{a, i}\n_{i\to a}(x_i),\quad\ \text{for all}\  a\in\A, i\in\N(a)
\end{split}
\end{equation}
with $\kappa_{a, i},\tau_{a, i}>0$  $(a\in\A, i\in\N(a))$. Plugging \eqref{eq:rescaledmessages} into \eqref{eq:fp2a} we obtain
the following fixed-point equations for the messages $\{\m_{a\to i}(x_i), \n_{i\to a}(x_i)\}$ $(a\in\A, i\in\N(a))$.
\begin{equation}\label{eq:fp2b}
\begin{cases}
\begin{split}
&\kappa_{a, i}\m_{a\to i}(x_i)\\
&=\omega_{a,i}\Big(\prod_{j\in \N(a)\setminus i}\tau_{a, j}\Big)\sum_{\xva\setminus x_i} f_a(\xva)\prod_{j\in \N(a)\setminus i}\n_{j\to a}(x_j)\\
&\tau_{a,i}\n_{i\to a}(x_i)\\
&=  \Big(\prod_{c\in \M(i)\setminus a}\kappa_{c, i}\Big)\prod_{c\in \M(i)\setminus a}\m_{c\to i}(x_i)
\end{split}
\end{cases}
\end{equation}
for all $a\in\A, i\in\N(a)$.
Now \eqref{eq:fp2b} is equivalent to \eqref{eq:fp2} if and only if
\begin{align}%
\tau_{a,i}&= \prod_{c\in \M(i)\setminus a}\kappa_{c, i},\quad\text{for all}\  a\in\A, i\in\N(a)\label{eq:condequiva}\\
\d_a
&=\frac{\omega_{a,i}\prod\limits_{j\in \N(a)\setminus i}\tau_{a, j}}{\kappa_{a, i}},\quad\text{for all}\   a\in\A, i\in\N(a)\label{eq:condequivb}
\end{align}
where the positive constants $\d_a$ $(a\in\A)$ are such that the beliefs $b_a$ $(a\in\A)$ in \eqref{stationarybethe} are normalized to one.
This normalization of the beliefs $b_a$ $(a\in\A)$ in \eqref{stationarybethe} gives
\begin{align}\label{eq:xxxxx}
\frac{1}{\d_a}
&=\sum\limits_{\xva}f_a(\xva)\prod\limits_{j\in \N(a)}\n_{j\to a}(x_j)\nonumber\\
&=\frac {\sum\limits_{\xva}f_a(\xva)\prod\limits_{j\in \N(a)}\nt_{j\to a}(x_j)}{\prod\limits_{j\in \N(a)}\tau_{a, j}}\nonumber\\
&=\frac{1}{\dt_a\prod\limits_{j\in \N(a)}\tau_{a, j}},\quad\text{for all}\  a\in\A
\end{align}
where we used \eqref{eq:rescaledmessages} in the second step and \eqref{eq:datilde} in the last step.
Combining \eqref{eq:condequiva}, \eqref{eq:condequivb}, and \eqref{eq:xxxxx} we obtain
\begin{align*}
\frac{1}{\dt_a}
&=\frac{\kappa_{a, i}\tau_{a,i}}{\omega_{a,i}}\\
&=\frac{g_i}{\omega_{a,i}},\quad\text{for all}\  a\in\A, i\in\N(a)
\end{align*}
with
\begin{align*}
g_i\triangleq \prod_{c\in \M(i)}\kappa_{c, i},\quad\text{for all}\  i\in\I.
\end{align*}

Now suppose that \eqref{eq:condbetter} is fulfilled. Setting
\begin{align*}
\kappa_{a, i}&=g_i^{\frac{1}{|\M(i)|}},\quad\text{for all}\   a\in\A, i\in\N(a)\\
\tau_{a, i}&=g_i^{1-\frac{1}{|\M(i)|}},\quad\text{for all}\   a\in\A, i\in\N(a)
\end{align*}
and reversing all the steps finishes the proof.

\subsection{Extension of the MF approximation to continuous random variables}\label{AppcontMF}
Suppose that $p_{\Xm}$ is a pdf  
of the vector of random variables $\Xm$. In this appendix, we assume that all integrals in the region-based free energy approximation are Lebesgue integrals and have finite values, which can be verified by inspection
of the factors $f_a$ ($a\in\A$) and the analytic expressions of the computed beliefs $b_i$ ($i\in\I$). An example where the MF approximation is applied to continuous random variables and combined with BP is discussed in Section \ref{Application}.

For each $i\in\I$ we can rewrite $F_\text{MF}$ in \eqref{eq:FMF} as
\begin{align*}
F_\text{MF}
=&
\,D(b_i\,||\,a_i)
+\sum_{j\in\I\setminus i}\int b_j(x_j) \ln b_j(x_j) \operatorname{d}\!x_j\nonumber\\
&-\sum_{a\in\A\setminus\M(i)}\int \ln f_a(\xva) \prod\limits_{j\in \N(a)} b_j(x_j)\operatorname{d}\!x_j\
\end{align*}
with
\begin{align*}
a_i(x_i)&
\triangleq
\exp\Big(\sum_{a\in\M(i)}\int \ln f_a(\xva) \prod\limits_{j\in \N(a)\setminus i} b_j(x_j)\operatorname{d}\!x_j\Big),\\
&\phantom{=}\ \text{for all}\  i\in\I.
\end{align*}
It follows from \cite[Th. 2.1]{ku78} that $D(b_i\,||\,a_i)$ is minimized subject to $\int b_i(x_i)\operatorname{d}\!x_i=1$ if and only if
\begin{align}\label{eq:cont}
b_i(x_i)&=\frac{a_i(x_i)}{\int a_i(x_i)\operatorname{d}\!x_i}
\end{align}
up to sets of Lebesgue measure zero. 
Formally, $b_i$ in \eqref{eq:cont} differs from $b_i$ in \eqref{eq:updateMF} by replacing sums  with Lebesgue integrals.

\subsection{Proof of Theorem \ref{Theoremcombined}}\label{Proofcombined}
The proof of Theorem \ref{Theoremcombined} is based on the ideas of the proof of \cite[Th.\,2]{yefrwe04}.
However, we will see that we get
a significant simplification by augmenting it with some of the arguments originally used in \cite{wajawi05} for Markov random fields and adopted to
factor graphs in \cite{wypesa11}. In particular, we shall make use of the following observation.
Recall the expression for $F_\text{BP,\,MF}$ in \eqref{Fcombined}
\begin{align}\label{Fcombined2}
\! F_\text{BP,\,MF}=&
\sum_{a\in\A_{\text{BP}}}\sum_{\xva}b_a(\xva)\ln\frac{b_a(\xva)}{f_a(\xva)}\nonumber\\
&-\sum_{a\in\A_{\text{MF}}}\sum_{\xva}\prod_{i\in \N(a)} b_i(x_i)\ln f_a(\xva)\nonumber\\
&-\sum_{i\in\I}(|\M_{\text{BP}}(i)|-1)\sum_{x_i}b_i(x_i)\ln b_i(x_i)
\end{align}
the marginalization constraints
\begin{align}\label{eq:marginalizationconstrains}
b_i(x_i)=\sum_{\xva\setminus x_i}b_a(\xva),\quad\text{for all}\  a\in\A_{\text{BP}}, i\in\N(a)
\end{align}
and the normalization constraints
\begin{align}\label{eq:normalizationconstrains}
\begin{split}
\sum_{x_i}b_i(x_i)&=1,\quad\text{for all}\  i\in\I_{\text{MF}}\setminus\I_{\text{BP}}\\
\sum_{\xva} b_a(\xva)&=1,\quad\text{for all}\  a\in\A_{\text{BP}}.
\end{split}
\end{align}
Using the marginalization constraints \eqref{eq:marginalizationconstrains},
we see that
\begin{align}\label{eq:trickW}
&\sum_{a\in\A_{\text{BP}}}\sum_{\xva}b_a(\xva)\ln \prod_{i\in \N(a)} b_i(x_i)\nonumber\\
&=\sum_{a\in\A_{\text{BP}}}\sum_{\xva}\sum_{i\in \N(a)}b_a(\xva)\ln  b_i(x_i)\nonumber\\
&=\sum_{a\in\A_{\text{BP}}}\sum_{i\in \N(a)}
\sum_{x_i}b_i(x_i)\ln  b_i(x_i)\nonumber\\
&=\sum_{i\in \I_\text{BP}}\sum_{a\in\M_\text{BP}(i)}
\sum_{x_i}b_i(x_i)\ln  b_i(x_i)\nonumber\\
&=\sum_{i\in \I_\text{BP}}|\M_\text{BP}(i)|\sum_{x_i}b_i(x_i)\ln  b_i(x_i).
\end{align}
Combining \eqref{eq:trickW} with \eqref{Fcombined2}, we further get
\begin{align}\label{Fcombined3}
\! F_\text{BP,\,MF}=&
-\sum_{a\in\A_{\text{BP}}}\sum_{\xva}b_a(\xva)\ln f_a(\xva)\nonumber\\
&-\sum_{a\in\A_{\text{MF}}}\sum_{\xva}\prod_{i\in \N(a)} b_i(x_i)\ln f_a(\xva)\nonumber\\
&+\sum_{i\in\I}\sum_{x_i}b_i(x_i)\ln b_i(x_i)\nonumber\\
&+\sum_{a\in\A_{\text{BP}}}I_a
\end{align}
with the mutual information \cite[p.\,19]{Cover91}
\begin{align*}
I_a&\triangleq \sum_{\xva}b_a(\xva)\ln\frac{b_a(\xva)}{\prod_{i\in \N(a)} b_i(x_i)},\quad\text{for all}\  a\in\A_\text{BP}.
\end{align*}
Next, we shall compute the stationary points of the Lagrangian
\begin{align}\label{Lcombined2}
L_\text{BP,\,MF}
=&F_\text{BP,\,MF}\nonumber\\
&
-\sum_{a\in\A_{\text{BP}}}\sum_{i\in \N(a)}\sum_{x_i}\lambda_{a,i}(x_i)
\Big(b_i(x_i)-\sum_{\xva\setminus x_i}b_a(\xva)\Big)\nonumber\\
&-\sum_{i\in\I_\text{MF}\setminus\I_\text{BP}}\gamma_i\Big( \sum_{x_i}b_i(x_i) -1\Big)\nonumber\\
&-\sum_{a\in\A_\text{BP}}\gamma_a\Big( \sum_{\xva}b_a(\xva) -1\Big)
\end{align}
using the expression for $F_\text{BP,\,MF}$ in \eqref{Fcombined3}. The particular form of $F_\text{BP,\,MF}$ in
\eqref{Fcombined3} is convenient because the marginalization constraints in \eqref{eq:marginalizationconstrains} imply that for all $i\in\I$ and
$a\in\A_\text{BP}$ we have
$\frac{\partial I_a}{\partial b_i(x_i)}=-\IND_{\M_\text{BP}(i)}(a)$. 
Setting the derivative of $L_\text{BP,\,MF}$ in \eqref{Lcombined2} with respect to $b_i(x_i)$ and $b_a(\xva)$ equal to zero for all
$i\in\I$ and $a\in\A_{\text{BP}}$, we  get the following fixed-point equations for the stationary points:
\begin{align}
\begin{split}\label{eq:stat1}
\ln b_i(x_i) =&
\sum\limits_{a\in \M_{\text{BP}}(i)}\lambda_{a,i}(x_i)\\
&+\sum\limits_{a\in \M_{\text{MF}}(i)}
\sum\limits_{\xva\setminus x_i}\prod\limits_{j\in \N(a)\setminus i} b_j(x_j) \ln f_a(\xva)\\
&+|\M_\text{BP}(i)|+\IND_{\I_\text{MF}\setminus\I_\text{BP}}(i)\gamma_i-1,\quad\text{for all}\  i\in\I\\
\ln b_a(\xva)=&\ln f_a(\xva) - \sum_{i\in\N(a)}\lambda_{a,i}(x_i) +\ln\Big(\prod_{i\in\N(a)}b_i(x_i)\Big)\\
& +\gamma_a-1,
\quad\text{for all}\  a\in\A_\text{BP}.
\end{split}
\end{align}
Setting
\begin{align}\label{eq:intmessages}
\begin{split}
 \m^{\text{BP}}_{a\to i}(x_i)\triangleq&\, \exp\Big(\lambda_{a,i}(x_i)+1-\frac{1}{|\M_\text{BP}(i)|}\Big),\\
 &\, \text{for all}\  a\in \A_\text{BP}, i\in\N(a)\\
 \m^{\text{MF}}_{a\to i}(x_i)\triangleq&\,
\exp\Biggl(\ \sum_{\xva\setminus x_i}\prod_{j\in \N(a)\setminus i}b_j(x_j)\ln f_a(\xva)\Biggr),\\
&\, \text{for all}\  a\in \A_\text{MF}, i\in\N(a)
\end{split}
\end{align}
we can rewrite \eqref{eq:stat1} as

\begin{align}
\begin{split}\label{eq:stat2}
b_i(x_i)
&=
\e_i
\prod\limits_{a\in \M_{\text{BP}}(i)} \m^{\text{BP}}_{a\to i}(x_i)
\prod\limits_{a\in \M_{\text{MF}}(i)}
 \m^{\text{MF}}_{a\to i}(x_i),\\
 &\phantom{=}\ \text{for all}\  i\in\I\\
b_a(\xva)
&
=
\d_a\, f_a(\xva)
\prod_{i\in\N(a)}
\frac{b_i(x_i)}{
\m^{\text{BP}}_{a\to i}(x_i)},\\
 &\phantom{=}\ \text{for all}\ a\in\A_{\text{BP}}
\end{split}
\end{align}
where
\begin{align*}
\e_i&\triangleq
\exp(\IND_{\I_\text{MF}\setminus\I_\text{BP}}(i)\gamma_i),\quad\text{for all}\ i\in\I\\
\d_a&\triangleq
\exp\Bigg(\gamma_a-1+\sum_{i\in\N(a)}\Big(1-\frac{1}{|\M_\text{BP}(i)|}\Big)\Bigg),\\
 &\phantom{=}\ \text{for all}\  a\in\A_\text{BP}
\end{align*}
are such that the normalization constraints in \eqref{eq:normalizationconstrains} are fulfilled.
Finally, we define
\begin{align}\label{eq:intnmessage}
 \n_{i\to a}(x_i)&\triangleq \e_i
\prod\limits_{c\in \M_{\text{BP}}(i)\setminus \{a\}} \m^{\text{BP}}_{c\to i}(x_i)
\prod\limits_{c\in \M_{\text{MF}}(i)} \m^{\text{MF}}_{c\to i}(x_i)
\end{align}
for all $a\in\A, i\in\N(a)$.
Plugging the expression for $\n_{i\to a}(x_i)$ in \eqref{eq:intnmessage} into the expression for $b_a(\xva)$ in \eqref{eq:stat2}
, we find that
\begin{align}
\begin{split}\label{eq:stat4}
b_i(x_i)&=
\e_i\prod\limits_{a\in \M_{\text{BP}}(i)} \m^{\text{BP}}_{a\to i}(x_i)
\prod\limits_{a\in \M_{\text{MF}}(i)} \m^{\text{MF}}_{a\to i}(x_i),\\
 &\phantom{=}\ \text{for all}\  i\in\I\\
b_a(\xva)&=\d_a\,
f_a(\xva)\prod\limits_{i\in \N(a)} \n_{i\to a}(x_i),\\
 &\phantom{=}\ \text{for all}\ a\in\A_{\text{BP}}.
\end{split}
\end{align}
Using the marginalization constraints in
\eqref{eq:marginalizationconstrains} in combination with \eqref{eq:stat4}  and noting that $\e_i=1$ for all $i\in\I_\text{BP}$
we further find that
\begin{align}\label{eq:xxx}
\n_{i\to a}(x_i) \m^{\text{BP}}_{a\to i}(x_i)
&=\prod\limits_{a\in \M_{\text{BP}}(i)} \m^{\text{BP}}_{a\to i}(x_i)\prod\limits_{a\in \M_{\text{MF}}(i)} \m^{\text{MF}}_{a\to i}(x_i)\nonumber\\
&= b_i(x_i)\nonumber\\
&=\sum_{\xva\setminus x_i}b_a(\xva)\nonumber\\
&=\d_a\sum_{\xva\setminus x_i}f_a(\xva)\prod\limits_{j\in \N(a)}  \n_{j\to a}(x_j)
\end{align}
for all $a\in\A_{\text{BP}}, i\in\N(a)$. Dividing both sides of \eqref{eq:xxx} by $ \n_{i\to a}(x_i)$ gives
\begin{align}\label{eq:xxx2}
 \m^{\text{BP}}_{a\to i}(x_i)&= \d_a\sum_{\xva\setminus x_i}f_a(\xva)\prod\limits_{j\in \N(a)\setminus i} \n_{j\to a}(x_j)
\end{align}
for all $a\in\A_{\text{BP}}, i\in\N(a)$. 
Noting that $\n_{j\to a}(x_j)=b_j(x_j)$ for all $a\in\A_\text{MF}$ and $j\in\N(a)$, we can write the messages
$m^{\text{MF}}_{a\to i}(x_i)$ in \eqref{eq:intmessages} as
\begin{align}\label{eq:mMF}
 \m^{\text{MF}}_{a\to i}(x_i)&=
\exp\Biggl(\ \sum_{\xva\setminus x_i}\prod_{j\in \N(a)\setminus i}
\n_{j\to a}(x_j)
\ln f_a(\xva)\Biggr)
\end{align}
for all $a\in\A_{\text{MF}}, i\in\N(a)$. 
Now \eqref{eq:intnmessage}, \eqref{eq:xxx2}, and \eqref{eq:mMF} are equivalent to \eqref{FPcombined} and \eqref{eq:stat4} is equivalent to
\eqref{bcombined}. This completes the proof that stationary points of the Lagrangian in \eqref{Lcombined} must be fixed-points with
positive beliefs fulfilling \eqref{bcombined}. Since all the steps are reversible, this also completes the proof of Theorem  \ref{Proofcombined}.

\subsection{Proof of Lemma \ref{theohard}}\label{proofhard}
We rewrite  $F_\text{BP,\,MF}$ in \eqref{Fcombined} as $F_\text{BP,\,MF}=F_1+F_2+F_3$ with

\begin{align*}
F_1&\triangleq\sum_{a\in\A_{\text{BP}}}D(b_a\ ||\ f_a) \\
F_2&\triangleq \sum_{a\in\A_{\text{MF}}}  D\Big( \prod_{i\in \N(a)} b_i\ ||\ f_a\Big)\\
F_3&\triangleq-\sum_{i\in\I}(|\M_{\text{BP}}(i)|+|\M_{\text{MF}}(i)|-1)\sum_{x_i}b_i(x_i)\ln b_i(x_i)
\end{align*}
and set
\begin{align*}
0<k_a\triangleq \sum_{\xva} f_a(\xva),\quad\text{for all}\  a\in\A.
\end{align*}
Then
\begin{align*}
F_1
=&\sum_{a\in\A_{\text{BP}}}D(b_a\, ||\, f_a/k_a)- \sum_{a\in\A_{\text{BP}}}\ln k_a \\
\geq&- \sum_{a\in\A_{\text{BP}}}\ln(k_a)\\
>&-\infty\\
F_2
=&\sum_{a\in\A_{\text{MF}}}D\Big( \prod_{i\in \N(a)} b_i\ ||\ f_a/k_a\Big)-\sum_{a\in\A_{\text{MF}}}\ln k_a\\
\geq& -\sum_{a\in\A_{\text{MF}}}\ln k_a\\
>&-\infty\\
F_3\geq&\ 0.
\end{align*}
This proves 1). Now $F_3<\infty$, \eqref{eq:faMF} implies that $F_2<\infty$, and \eqref{eq:faBP} implies that $F_1<\infty$ if and only if \eqref{eq:removestate1} if fulfilled, which proves 2).

Suppose that we have fixed all $b_a(\xvba)$ ($a\in\A_{\text{BP}}$) from \eqref{eq:removestate1}.
Then the  analysis for the remaining $b_i(x_i)$ $(i\in\I)$  and  $b_a(\xva)$ excluding all
$\xvba$ from \eqref{eq:removestate1} $(a\in\A_\text{BP})$ is the same as in the proof of Theorem \ref{Theoremcombined} and the resulting fixed-point equations are identical to \eqref{bcombined} and \eqref{FPcombined}
excluding all $\xvba$ from \eqref{eq:removestate1} and vice versa, which proves 3). We can reintroduce the realizations $\xvba$ with $f_a(\xvba)=0$ ($a\in\A_\text{BP}$) from \eqref{eq:removestate1} in  \eqref{FPcombined} because they do not contribute to the  message passing update equations, as can be seen immediately from the definition of the messages $m^{\text{BP}}_{a\to i}(x_i)$ $(a\in\A_\text{BP}, i\in\N(a))$ in \eqref{FPcombined}. The same argument implies that \eqref{eq:removestate1} is a special case of the first equation in \eqref{bcombined}, which proves 4) and, therefore, finishes the proof of Lemma \ref{theohard}.

\subsection{Proof of convergence}\label{Aconvergence}
In order to finish the proof of convergence for the algorithm presented in Subsection \ref{Rconvergence}, we need to show that running the forward/backward algorithm in the BP part in step 2) of Algorithm \ref{Algorithm} cannot increase the region-based free energy approximation $F_{\text{BP,\,MF}}$ in \eqref{Fcombined}.
To this end we analyze the factorization
\begin{align}\label{eq:facf}
p(\xv_\text{BP})&\propto \prod_{a\in \A_{\text{BP}}}f_a(\xva)
\prod_{i\in\I_\text{BP}\cap\I_\text{MF}}\prod_{b\in\M_{\text{MF}}(i)}m^{\text{MF}}_{b\to i}(x_i)
\end{align}
with $\xv_\text{BP}\triangleq(x_i\mid i\in \I_{\text{BP}})^{\operatorname{T}}$. The factorization in \eqref{eq:facf} is the product of the factorization of the BP part in
\eqref{eq:factorcombined} and the
incoming messages from the MF part. 
The Bethe free energy \eqref{bethefree} corresponding to the factorization in \eqref{eq:facf} is
\begin{align}
F_\text{BP}=&
\sum_{a\in\A_{\text{BP}}}\sum_{\xva}b_a(\xva)\ln\frac{b_a(\xva)}{f_a(\xva)}\nonumber\\
&+\sum_{i\in\I_\text{BP}\cap\I_\text{MF}}\sum_{a\in\M_{\text{MF}}(i)}\sum_{x_i}b_i(x_i)\ln \frac{b_i(x_i)}{m^{\text{MF}}_{a\to i}(x_i)}
\nonumber\\
&-\sum_{i\in\I_\text{BP}}(|\M_{\text{BP}}(i)|+|\M_{\text{MF}}(i)|-1)\sum_{x_i}b_i(x_i)\ln b_i(x_i)\nonumber\\
=&\sum_{a\in\A_{\text{BP}}}\sum_{\xva}b_a(\xva)\ln\frac{b_a(\xva)}{f_a(\xva)}\nonumber\\
&-\sum_{i\in\I_\text{BP}\cap\I_\text{MF}}\sum_{a\in\M_{\text{MF}}(i)}\sum_{x_i}b_i(x_i)\ln m^{\text{MF}}_{a\to i}(x_i)
\nonumber\\
&-\sum_{i\in\I_\text{BP}}(|\M_{\text{BP}}(i)|-1)\sum_{x_i}b_i(x_i)\ln b_i(x_i).
\label{eq:Ff}
\end{align}
We now show that minimizing $F_\text{BP}$ in \eqref{eq:Ff}
is equivalent to minimizing
$F_\text{BP,\,MF}$ in \eqref{Fcombined} with respect to $b_a$ and $b_i$ for all $a\in\A_\text{BP}$ and
$i\in\I_\text{BP}$. Obvioulsy,
\begin{align*}
\frac{\partial F_\text{BP,\,MF}}{\partial b_i(x_i)}&=\frac{\partial F_\text{BP}}{\partial b_i(x_i)},\quad \text{for all}\ i\in \I_\text{BP}\setminus \I_\text{MF}
\end{align*}
and
\begin{align*}
\frac{\partial F_\text{BP,\,MF}}{\partial b_a(\xva)}&=\frac{\partial F_\text{BP}}{\partial b_a(\xva)},\quad\text{for all}\  a\in \A_\text{BP}.
\end{align*}
This follows from the fact that $F_\text{BP,\,MF}$ differs from $F_\text{BP}$ by terms that depend only on $b_i$ with
$i\in \I_\text{MF}$. Now suppose that $i\in \I_\text{BP}\cap\I_\text{MF}$. In this case, we find that
\begin{align}\label{eq:partialconv1}
\frac{\partial F_\text{BP,\,MF}}{\partial b_i(x_i)}
&=(1-|\M_{\text{BP}}(i)|)(\ln b_i(x_i)+1)\nonumber\\
&\phantom{=}-\sum\limits_{a\in \M_{\text{MF}}(i)}\sum\limits_{\xva\setminus x_i}\prod\limits_{j\in \N(a)\setminus i} b_j(x_j) \ln f_a(\xva)
\end{align}
and
\begin{align}\label{eq:partialconv2}
\frac{\partial F_\text{BP}}{\partial b_i(x_i)}
&=(1-|\M_{\text{BP}}(i)|)(\ln b_i(x_i)+1)
-\!\!\!\!\!\!\sum_{a\in\M_{\text{MF}}(i)}\ln m^{\text{MF}}_{a\to i}(x_i).
\end{align}
From   \eqref{FPcombined} we see that
\begin{align}\label{eq:mproof}
m^{\text{MF}}_{a\to i}(x_i)&=
\exp\Biggl(\ \sum_{\xva\setminus x_i}\prod_{j\in \N(a)\setminus i}
\n_{j\to a}(x_j)
\ln f_a(\xva)\Biggr)
\end{align}
for all $a\in\M_\text{MF}(i)$. Note that, according to step 2) in Algorithm \ref{Algorithm}, the messages $m^{\text{MF}}_{a\to i}(x_i)$ in \eqref{eq:mproof} are {\em fixed inputs} for the BP part. Therefore, we are not allowed to plug the expressions for the messages $m^{\text{MF}}_{a\to i}(x_i)$ in \eqref{eq:mproof} into  \eqref{eq:partialconv2} in general. However, since $a\in\A_\text{MF}$ and $i\in \I_\text{BP}\cap\I_\text{MF}$, condition \eqref{eq:condconv} implies that
$\N(a)\setminus i\subseteq \I_\text{MF}\setminus \I_\text{BP}$ and guarantees that
\begin{align*}
\n_{j\to a}(x_j)
&=b_j(x_j)
\end{align*}
is  constant in step 2) of Algorithm \ref{Algorithm} for all $j\in \N(a)\setminus i\subseteq \I_\text{MF}\setminus \I_\text{BP}$.
Therefore, we are indeed allowed to plug the expressions of the messages $m^{\text{MF}}_{a\to i}(x_i)$ in \eqref{eq:mproof} into  \eqref{eq:partialconv2} and  finally see that also
\begin{align*}
\frac{\partial F_\text{BP,\,MF}}{\partial b_i(x_i)}&=\frac{\partial F_\text{BP}}{\partial b_i(x_i)},\quad \text{for all}\ i\in \I_\text{BP}\cap \I_\text{MF}.
\end{align*}
Hence, minimizing
 $F_\text{BP}$ in \eqref{eq:Ff} is equivalent to minimizing
$F_\text{BP,\,MF}$ in \eqref{Fcombined}.

By assumption, the factor graph in the BP part has a tree structure.  Therefore, \cite[Prop.\,3]{yefrwe04} implies that
\begin{enumerate}
\item $F_\text{BP}\geq 0$;
\item
$F_\text{BP} = 0$ if and only if the beliefs $\{b_i,b_a\}$  in \eqref{eq:Ff} are the marginals of the factorization in \eqref{eq:facf}.
\end{enumerate}
Hence, for $b_j$ fixed with $j\in\I_\text{MF}\setminus\I_\text{BP}$, we see that $F_\text{BP,\,MF}$ in \eqref{Fcombined} is minimized by the marginals of the factorization in \eqref{eq:facf}.

It remains to show that running the forward/backword algorithm in the BP part as described in step 2) in Algorithm
\ref{Algorithm} indeed computes the marginals of the factorization in \eqref{eq:facf}.
Applying Theorem \ref{the1} to the factorization in  \eqref{eq:facf} yields the message passing fixed-point equations
\begin{equation}
\label{FPconvergence}
\begin{cases}
\begin{split}
\n_{i\to a}(x_i)&=
\prod\limits_{c\in \M_{\text{BP}}(i)\setminus a}\m^{\text{BP}}_{c\to i}(x_i)
\prod\limits_{c\in \M_{\text{MF}}(i)}\m^{\text{MF}}_{c\to i}(x_i)\\
\m^{\text{BP}}_{a\to i}(x_i)&=\d_a
\sum\limits_{\xva\setminus x_i} f_a(\xva) \prod\limits_{j\in \N(a)\setminus i} \n_{j\to a}(x_j)\end{split}
\end{cases}
\end{equation}
for all $a\in \A_\text{BP}, i\in\N(a)$. The message passing fixed-point equations in \eqref{FPconvergence} are the same as the message passing fixed-point equations for the BP part in
 \eqref{FPcombined} with fixed-input messages $m^{\text{MF}}_{a\to i}(x_i)$ for all
 $i\in\I_{\text{BP}}\cap\I_{\text{MF}} $ and $a\in\M_{\text{MF}}(i)$. Hence, running the forward/backward algorithm in the BP part indeed computes the marginals of the factorization in  \eqref{eq:facf} and Algorithm \ref{Algorithm} is guaranteed to converge.
\subsection{Product of Gaussian distributions}\label{Agaussian}
\begin{lemma}\label{lemmagaussian}
Let
\begin{align*}
p_i(\xv)&=\text{CN}(\xv;\muv_i,\Lam_i^{-1}),\quad\text{for all}\ i\in[1:N].
\end{align*}
Then
\begin{align*}
\prod_{i\in[1:N]}p_i(\xv)&\propto\text{CN}(\xv;\muv,\Lam^{-1})
\end{align*}
with
\begin{align*}
\muv&\triangleq\sum_{i\in[1:N]}\Lam^{-1}\Lam_i\muv_i\\\
\Lam&\triangleq\sum_{i\in[1:N]}\Lam_i.
\end{align*}
\end{lemma}
\proof Follows from direct computation.\endproof


\begin{IEEEbiographynophoto}{Erwin Riegler}
(M'07) received the 
Dipl-Ing. degree in Technical Physics (with distinction) in 2001 and the 
Dr. techn. degree in Technical Physics (with distinction) in 2004 from Vienna University of Technology. 

He was a visiting researcher at the 
Max Planck Institute for Mathematics in the Sciences in Leipzig, Germany (Sep. 2004 - Feb. 2005),  
the Communication Theory Group at ETH Z\"urich, Switzerland (Sep. 2010 - Feb. 2011), 
and the Department of Electrical and Computer Engineering at The Ohio State University in Columbus, Ohio (Mar. 2012). 
From 2005 - 2006, he was a post-doc at the Institute for Analysis and Scientific Computing, Vienna University of Technology. 
From 2007 - 2010, he was a senior researcher at the Telecommunications Research Center Vienna (FTW). 
Since 2010, he has been a post-doc at the Institute of Telecommunications at Vienna University of Technology. 

His research interests include noncoherent communications, machine learning, interference management, large system analysis,
 and transceiver design.
\end{IEEEbiographynophoto}

\begin{IEEEbiographynophoto}{Gunvor Elisabeth Kirkelund}
received her master degree (cum laude) in Wireless Communication from Aalborg University, Denmark, in 2008.

Since 2008 she has been pursuing a Ph.D. degree at the Section Navigation and Communications, Department of Electronic Systems, Aalborg University. Her research interests lie within the field of statistical signal processing, message-passing techniques and design of iterative algorithms for wireless receivers.
\end{IEEEbiographynophoto}

\begin{IEEEbiographynophoto}{Carles Navarro Manch\'on}
received the degree in telecommunications engineering from the University Miguel Hern\'andez of Elx, Spain, in 2006 and the PhD degree in wireless communications from Aalborg University, Denmark, in 2011.

Since 2006 he has been with Aalborg University, where he is currently an Assistant Professor in the Section Navigation and Communications, Department of Electronic Systems. His research interests lie within the area of statistical signal processing for wireless communications, including joint channel estimation and decoding, distributed signal processing for cooperative communications and estimation and reconstruction of sparse signals.
\end{IEEEbiographynophoto}

\begin{IEEEbiographynophoto}{Mihai-Alin Badiu}
received the Dipl.-Ing. degree in electrical engineering in June 2008 and the Master degree in telecommunications in June 2010 from the Technical University of Cluj-Napoca, Romania. Since Oct. 2009, he is pursuing a Ph.D. degree at the Communications Department of the same university. 
	
From Nov. 2008 to May 2010, he was a Research Assistant at the Technical University of Cluj-Napoca, Romania. In 2011 he spent an 8-month stay as a Visiting Researcher with the Section Navigation and Communications of the Department of Electronic Systems, Aalborg University, Denmark. Since Feb. 2012 he is affiliated with this Section as a Research Assistant. His research interests include advanced wireless receiver design based on message-passing algorithms and cooperative communications.
\end{IEEEbiographynophoto}

\begin{IEEEbiographynophoto}{Bernard H. Fleury}
(M'97--SM'99) received the Diploma in electrical engineering and mathematics in 1978 and 1990, respectively, and the Ph.D. degree in electrical engineering from the Swiss Federal Institute of Technology Zurich (ETHZ), Switzerland, in 1990. 

Since 1997, he has been with the Department of Electronic Systems, Aalborg University, Denmark, as a Professor of communication theory. He is the Head of the Section Navigation and Communications, which is one of the eleven laboratories of this Department. From 2006 to 2009, he was a Key Researcher with the Telecommunications Research Center Vienna (FTW), Austria. During 1978--1985 and 1992--1996, he was a Teaching Assistant and a Senior Research Associate, respectively, with the Communication Technology Laboratory, ETHZ. Between 1988 and 1992, he was a Research Assistant with the Statistical Seminar at ETHZ.

Prof. Fleury's research interests cover numerous aspects within communication theory, signal processing, and machine learning Ð mainly for wireless communications. His current scientific activities include stochastic modelling and estimation of the radio channel especially for MIMO systems operating in harsh conditions, iterative message-passing processing with focus on the design of efficient feasible architectures for wireless receivers, localization techniques in wireless terrestrial systems, and radar signal processing. He has authored and co-authored more than 120 publications in these areas. He has developed with his staff a high-resolution method for the estimation of radio channel parameters that has found a wide application and has inspired similar estimation techniques both in academia and in industry.
\end{IEEEbiographynophoto}

\end{document}